# Electron Channelling Contrast SEM Imaging of Twist Domains in Transition Metal Dichalcogenide Heterostructures

*Evan Tillotson[1,3], Dr. James McHugh[2,3], Dr. James Howarth[3], Dr. Teruo Hashimoto[1], Dr. Nick Clark[1,3], Dr. Astrid Weston[2,3], Dr. Vladimir Enaldiev[2,3], Samuel Sullivan-Allsop[1,3], William Thornley[1,3], Dr. Wendong Wang[2,3], Dr. Matthew Lindley[1], Dr. Andrew Pollard[4], Prof. Vladimir Falko[2,3^], Prof. Roman Gorbachev[2,3#] and Prof. Sarah J. Haigh[1,3*].*

1. Department of Materials, University of Manchester, Manchester M13 9PL, UK
2. Department of Physics and Astronomy, University of Manchester, Manchester M13 9PL, UK
3. National Graphene Institute, University of Manchester, Manchester M13 9PL, UK
4. National Physical Laboratory, Hampton Rd, Teddington, TW11 0LW, UK

*sarah.haigh@manchester.ac.uk; #roman@manchester.ac.uk; ^Vladimir.falko@manchester.ac.uk.

**Abstract**

Twisted 2D material heterostructures provide an exciting platform for investigating new fundamental physical phenomena. Many of the most interesting behaviours emerge at small twist angles, where the materials reconstruct to form areas of perfectly stacked crystal separated by partial dislocations. However, understanding the properties of these systems is often impossible without correlative imaging of their local reconstructed domain architecture, which exhibits random variations due to disorder and contamination. Here we demonstrate a simple and widely accessible route to visualise domains in as-produced twisted transition metal dichalcogenide (TMD) heterostructures using electron channelling contrast imaging (ECCI) in the scanning electron microscope (SEM). This non-destructive approach is compatible with conventional substrates and allows domains to be visualised even when sealed beneath an encapsulation layer. Complementary theoretical calculations reveal how a combination of elastic and inelastic scattering leads to contrast inversions at specified detector scattering angles and sample tilts. We demonstrate that optimal domain contrast is therefore achieved by maximising signal collection while avoiding contrast inversion conditions.

**Introduction**

Van der Waals (vdW) heterostructures offer a new degree of freedom for materials design through the ability to control the twist angle between adjacent atomic layers. When 2D crystals in contact have a similar lattice constant and/or a small angular misalignment, a moiré superlattice forms at their interface.[1] When the superlattice repeat period is long enough, the system can spontaneously reconstruct, forming pristine crystalline domains separated by narrow dislocations. This long-range periodicity of twist domains can dramatically affect the (opto)electronic properties of the system,[2] and has become a topic of active research referred to as "twistronics".[3–5] The ability to tune the electronic structure through reconstructed twist domains has enabled studies of correlation effects in flat electronic bands,[6,7] revealing exciting physical phenomena such as superconductivity in graphene,[8] exciton trapping,[9] resonant excitonic states[10,11] and excitonic density waves.[12] Unfortunately, both top-down mechanical assembly and bottom-up chemical synthesis of vdW heterostructures often result in large variability in the local twist angle and therefore large variability in the size of the reconstructed twist domains, even within single crystal samples.[13] In the case of vdW assembly from exfoliated crystals, this inhomogeneity originates from random strain fields acquired during the nanofabrication process and manifests as approximately 0.1° twist angle disorder.[14] This is problematic because the properties of reconstructed twisted 2D structures are sensitive to small local twist angle variation,[15] making the correlation of lattice distortion and electronic properties necessary for the future progress of twistronics. [16]

To date, various techniques have been utilised for mapping the reconstructed domain networks in twisted heterostructures. Scanning tunnelling microscopy (STM) allows high spatial resolution imaging *via* probing of the moiré-induced variations in the local electronic structure.[17,18] Other, less demanding, scanning probe techniques such as piezoresponse force microscopy allow sub-5-nm visualisation.[19] Aberration-corrected low energy electron microscopy (LEEM) can visualise the average atomic structure of domains and strain information.[20] However, these techniques require the twisted interface to be at the surface. As the majority of 2D material devices require encapsulation,[21] typically in hexagonal boron nitride (hBN), either due to their air sensitivity, to increase their (opto)electronic performance, or for electrostatic gating, an alternative non-destructive imaging approach is required.

Transmission electron microscopy (TEM) allows imaging of the microstructure of twisted superlattices down to the atomic scale,[22,23] but the technique requires suspended samples, which are often incompatible with property measurements, making it impossible to discern the dependence of the (opto)electronic properties on the local atomic configuration. Scanning near-field optical microscopy (SNOM) allows imaging of domain networks where they provide sufficiently strong confined light-matter interactions (polaritons), for example in twisted graphene through detecting plasmon reflections at domain boundaries[24] or exploiting phonons in hBN.[25] Nonetheless, this is a relatively specialised technique which requires artificial doping of the heterostructure, so is not generally applicable to unmodified optoelectronic device structures.[24–27]

Scanning electron microscopy (SEM) imaging has many advantages for imaging twist domains in 2D heterostructures, being a widely available, non-destructive, and surface sensitive technique compatible with most electronic device configurations. More specifically, electron channelling contrast, where an enhanced electron yield occurs at specific lattice orientations, has been widely applied for SEM imaging of crystal grain structures and dislocations in bulk materials.[28,29] While SEM electron channelling contrast imaging (ECCI) has rarely been employed to visualise twisted 2D materials,[2,30,31] a general understanding of the electron scattering processes arising from reconstructed 2D lattices is currently absent.

In this work we conduct a comprehensive SEM ECCI analysis of twisted TMD heterostructures. We analyse elastic and inelastic scattering contributions in both secondary electron (SE) and backscattered electron (BSE) detector signals and establish a complementary theoretical model to support our observations. We demonstrate the robustness of BSE ECCI imaging of reconstructed domains, being insensitive to stage rotation angle and which does not require sample tilting. Finally, we study the effects of sample thickness, choice of substrate, and hBN encapsulation. This work demonstrates the SEM ECCI method to be a widely accessible and practical imaging solution for the twistronics community.

**Results**

When two TMD monolayers are aligned close to parallel orientation of the unit cell (i.e., with a small angular mismatch) the lattice reconstruction results in the emergence of tessellated triangular domains (see **Figure 1a**).[32] The two alternating domain types are both rhombohedral (3R) but are (0001) mirror symmetry equivalent, with one denoted MX' (metal atoms in the top layer located above the chalcogen atoms in the lower layer) and the other XM' (vice versa). These stacking domains are separated by several-nanometre-wide domain walls which constitute a partial dislocation, illustrated schematically in **Figure 1b**. In all samples studied, mechanical transfer methods used during fabrication resulted in a noticeable variation in both domain periodicity and size due to the presence of random local strain and pockets of interlayer contamination.[33] Examples include wrinkles (white or dark lines) and contamination pockets (dark areas) which locally distort the triangular domain network. Both of these effects can be seen in the SEM BSE channelling contrast image in **Figure 1a**, acquired with an energy selective backscatter (EsB) electron detector. Here, an $MoS_2$ twisted bilayer was assembled using the tear-and-stack approach[34] and subsequently placed on a relatively thick (~50 nm) graphite crystal which had been exfoliated onto an oxidised silicon wafer.

*Domain contrast in suspended samples*

To deconvolute the complex picture of electron scattering and detection, we first focus our attention on freely suspended TMD bilayers where the substrate-induced scattering is absent. To fabricate these samples, twisted bilayers have been deposited onto few-nm thick hBN membranes with pre-drilled apertures and then placed over the holes in silicon nitride TEM grids (see methods and **SI Section 1** for full fabrication details). For our analysis, an archetypal SEM system (Zeiss Merlin Gemini II SEM) is used. Like most modern SEMs, this instrument is fitted with a variety of electron detectors both above and below the sample (**Figure 1c**). The most common detector in SEM imaging is the conventional Everhart-Thornley detector (ETD). This has a positive bias and is positioned at a relatively large distance from the scanned electron beam. ETD images are often termed 'secondary electron'

(SE) images but could more accurately be referred to as low energy electron images, since they will include information from any low energy electrons emitted from the sample surface. The maximum energy of these collected electrons varies with detector bias but is typically <100 eV.[35] The other common SEM imaging mode is referred to as backscattered electron (BSE) imaging, and is achieved using an annular detector placed directly above the sample. This type of detector uses a negative applied bias, which serves to repel electrons with lower energy, ensuring only high energy electrons emitted from the sample contribute to the signal. The Zeiss Merlin Gemini II SEM has two such detectors: the annular in-lens (InL) and Energy selective Backscattered (EsB) detectors, being distinguished by their different angular ranges relative to the incident electron probe. The SEM in this study also has a segmented scanning transmission electron microscopy (STEM) detector below the sample, which allows electron transparent regions to be studied in transmission mode. The detector collects electrons of any energy, and is split into 4 concentric annular segments with STEM1 being the central circle, and STEM2, STEM4 and STEM5 ring detectors with increasing radii as illustrated in **Figure 1c**.

**Figure 1d** shows simultaneously acquired SEM images for a freely suspended sample region consisting of a $MoS_2$ twisted bilayer with incident beam direction close to [0001], where reconstructed moiré superlattice domains are clearly visible as triangular regions with alternating contrast. We use an accelerating voltage of 1500 V and a compromise working distance of ~6 mm (we discuss the reasons for these settings later in the manuscript). To quantitatively compare the domain contrast for different detectors we use the Michaelson contrast, calculated from measurements of the average intensity in the image of XM' and MX' domains, $I_{XM'}$ and $I_{MX'}$ respectively, and defined as $(I_{XM'}-I_{MX'})/(I_{XM'}+I_{MX'})$. Our previous work allows us to determine that for these imaging conditions EsB images have brighter XM' domains and darker MX' domains, based on their response to an applied electric field.[32] Here we will define this as positive contrast, $I_{XM'}-I_{MX'}>0$, while the STEM2 detector image shows negative contrast, $I_{XM'}-I_{MX'}<0$. For further details of the method used for quantifying the domain contrast from the SEM images see **SI Section 2** and **Figures S2.1-3**.

In **Figure 1d**, the greatest domain contrast is observed for the EsB detector (Michaelson contrast of 21%), followed by the InL detector (20% contrast). The ETD image is not shown as it had insufficient signal to see any contrast in the freely suspended sample. Out of all the STEM detector segments, STEM2 and STEM5 showed the greatest domain contrast (-12.4% and 11.7%, respectively). We now seek to understand the origin of the surprising contrast inversion, whereby in the STEM2 the bright and dark domains are reversed compared to the EsB, InL and STEM5 images. Flipping the specimen upside-down causes the domain contrast of all images to invert as expected (since MX' domains then become XM' when inverted and vice versa, see **Figure S2.5**).

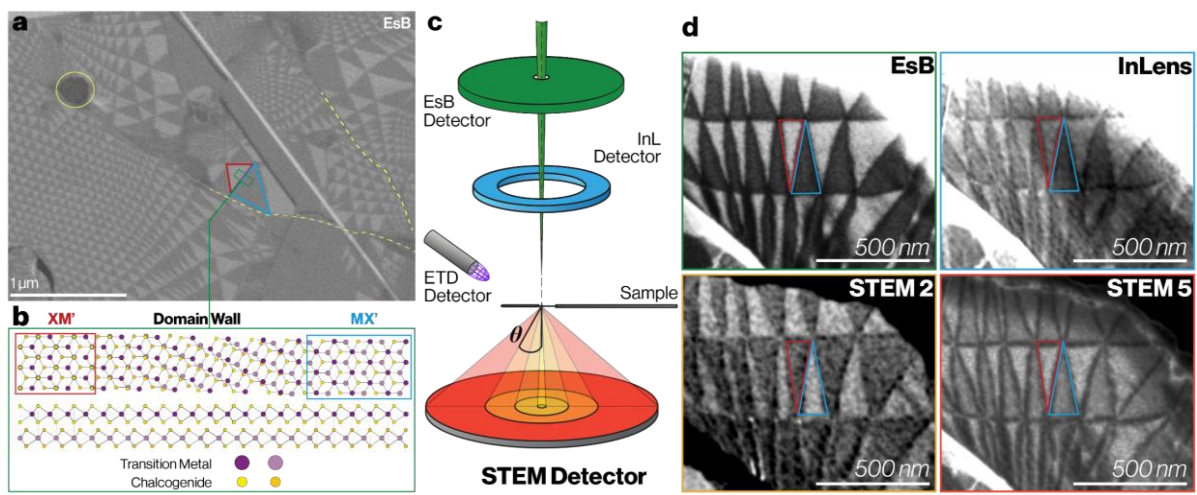

**Figure 1**. **SEM image of reconstructed domains in parallel stacked twisted TMD bilayer**. **a**) SEM EsB detector image showing a typical twisted 3R bilayer of $MoS_2$ on a graphite/$SiO_2$ substrate with visible disorder in the reconstructed domain network. Contamination pockets and wrinkles are shown *via* yellow circles and dashed lines, respectively. **b**) Schematic illustrating the atomic structure within reconstructed 3R type domains (XM' and MX' stacked) and at the domain wall for parallel, marginally twisted TMD crystals. Top: plan view along [0001]. Bottom: cross section viewed along [2-1-10]. In (**a**) examples of XM' and MX' stacked domains are highlighted by the red and blue triangles, respectively, with a legend depicting each atom for the illustration in **b**). **c**) Schematic showing the location of different detectors in the SEM relative to the sample, and the scattering angle θ. The STEM detector has 4 segments (STEM1, STEM2, STEM4, and STEM5) each having a different range of inner, outer collection angles, which can be slightly modified by changing the distance between the detector and the-sample. **d**) Comparing simultaneously acquired images of twisted superlattice domains in a suspended $MoS_2$ bilayer acquired with the EsB, InL, STEM2 (low angle transmitted electron) and STEM5 (high angle transmitted electron) detectors. Positive domain contrast is defined as where $I_{XM'}-I_{MX'}>0$ (red MX' domains brighter as in EsB, InLens and STEM5 images) and negative domain contrast is where $I_{XM'}-I_{MX'}<0$, (blue XM' domains brighter, as in STEM2 image).

First, we consider the domain contrast inversion seen in **Figure 1d** when comparing the transmitted (STEM) detector signal at low angle (STEM2) and high angle (STEM5), to shed light on the underlying electron scattering processes. To measure the domain contrast as a function of scattering angle, $\theta$, (see **Figure 1c**) we vary the annular range of the scattering angle, $\Delta\theta$, collected by each STEM detector segment by adjusting the sample-detector distance. The results are presented in **Figure 2d,** where each horizontal bar shows the integral contrast value for electrons scattered within an interval between $\theta_1$ and $\theta_2$, as denoted by the horizontal position of the bar. The greatest contrast is seen for collection angles in the range 4-12° (negative contrast) and the range 24-50° (positive contrast).

To understand the unusual contrast behaviour, and specifically the zero-contrast inversion point at ~20°, we have performed theoretical calculations of both elastic and inelastic scattering contributions to domain contrast. We first consider purely elastic scattering, during which the transmitted electrons undergo Bragg reflections. The resulting Bragg peaks, g, for XM' and MX' stacked domains can be expected to have different intensities, referred to as $I_{XM}(g)$ and $I_{MX}(g)$ respectively. The relative magnitude of $I_{XM}(g)$ and $I_{MX}(g)$ for each Bragg reflection ($I_{XM} - I_{MX}$) varies as a function of the incident electron beam energy, $E$ (see **Figure 2a**). At the experimental beam energy of 1500 eV, the first $g = \{\bar{1}\ 1\ 0\ 0\}$ Bragg peak occurs at $\theta \approx 7°$ with our calculations predicting that this will give strong negative domain contrast (see **Figure 2b**), which clearly correlates to the negative domain contrast values observed for the STEM2 detector with an angular range of $1.2° < \theta < 16°$. However, higher order Bragg reflections are also predicted to give negative domain contrast, so elastic scattering alone cannot explain the observed contrast inversion for $\theta \approx 20°$. That said, sign reversal is predicted for elastic contributions to STEM images at lower or higher incident beam energies, for example **Figure 2a** shows that at an incident beam energy of 1000 eV the Bragg peak at ~8° is negative but the peak at ~30° is positive (see **SI Section 3.1** for a full discussion and details of numerical calculations).

As contrast inversion cannot be explained by elastic scattering only, we must therefore consider the contribution from the electrons that are inelastically scattered. A semi-classical model of channelling contrast was employed, where generation and attenuation of secondary electrons are both proportional to electron density. Secondary electrons are considered to be generated by the primary beam, proportional to the local electron density of an $MoS_2$ bilayer, $p_{gen} \propto \rho(r)$ and the corresponding electron-electron cross section for a primary beam of energy $E_0$. Inelastic scattering of the primary beam transfers energy, $\Delta E$, to lattice electrons in the sample. The probability for an electron to escape as a result of this process is then taken as the integral along the escape path, with a cross-section corresponding to the transferred energy, $\Delta E$. Cross-sections were taken to reproduce the empirical TPP-2M mean-free paths,[36,37] which largely accounts for low-energy valence band scattering (plasmon and interband transitions) and does not fully incorporate large energy transfer and high angle deflections (see **SI Section 3.2** for full model details and **SI Section 3.2.1** for discussion of inelastic cross sections).

Using this model, the inelastic scattering behaviour variation is computed as a function of scattering angle, $\theta$, along a given in-plane crystallographic direction (zigzag, $\phi = 60^o$, or armchair $\phi = 30^o$) as shown in **SI Figures S3.11** and **S3.12**, respectively, for an incident energy of 1500 eV and for generated electrons with $\Delta E$=100 eV. The radially averaged sum is plotted in **Figure 2c** and shows a trough corresponding to maximum negative domain contrast at ~$16^o$, with peaks corresponding to maximum positive domain contrast at ~$22^o$ and ~$30^o$. It is therefore possible to reproduce a good match to the experimentally observed STEM detector contrast behaviour for scattering angles up to θ~35° by combining elastic and inelastic scattering contributions with an empirically determined weighting, illustrated by the black dots in **Figure 2d**. The annular dependence of secondary electrons with higher energy losses also shows oscillatory behaviour and is in phase with the $\Delta E$=100 eV data as shown in **Figure S3.15**. The discrepancy between the experiment and theory at higher scattering angles can be understood as a result of plural scattering of electrons or due to large energy loss associated with ionisation of core electrons (binding energies, 200–500 eV), neither of which were considered in this model (as discussed in **SI Section 3.2.4**).

Similar modelling can also explain the features observed in the reflected signals for suspended samples. The elastic signal reflected in the suspended bilayer (90°< $\theta$ <180°) is four orders of magnitude smaller than the transmitted elastic signal so it's contribution can be ignored (see **SI Figure S3.7**). **Figure S3.13** shows the low energy secondary electron signal for suspended samples as a function of scattering angle, $\theta$, which oscillates in a similar manner to the transmitted signal and so provides either positive or negative domain contrast depending on the angular range of the collected electrons. The EsB and InL both collect inelastically scattered electrons close to perfect reflection in the range 160°< $\theta$ <179° and 150°< $\theta$ <160° for the EsB and InL, respectively. In both angular ranges the domain contrast of the inelastic electrons is predicted to be positive; in agreement with experimental observations, although we note that for a full treatment, high energy inelastic scattering by core electrons should be included in the calculations. As the InL detector always gave similar images to the EsB detector, but with slightly lower contrast, from this point we focus on comparing ETD and EsB detector images.

The absence of visible contrast in the ETD image is assigned to the very low total signal collected on the ETD. This can be assigned to a geometric effect, as the ETD collection efficiency is highest for surfaces oriented parallel to the electron beam and weakest for surfaces perpendicular to the beam;[38] the ETD will therefore have poor collection efficiency for a flat suspended bilayer in the absence of sample tilt. Furthermore, the number of secondary electrons generated in such a thin suspended sample (1.4nm) will be comparatively small. Finally, the small escape depth of the low energy electrons collected by the ETD means these are unlikely to escape through both layers of the sample, which is required for channelling differences to create domain contrast, reducing the signal further. For example, 50 eV electrons have an escape depth of 0.48 nm, which is less than 1 layer of $MoS_2$ (see **SI Figure S3.9** for our calculations of the inelastic mean-free paths for $MoS_2$ as a function of electron energy).[39] While these calculations have been performed for MoS$_2$ bilayers, we find similar contrast inversion behaviour for WS$_2$ bilayers, although the Michaelson contrast values for WS$_2$ are approximately 2-3 times higher than those of MoS$_2$, which we assign to the higher atomic number promoting greater inelastic scattering.

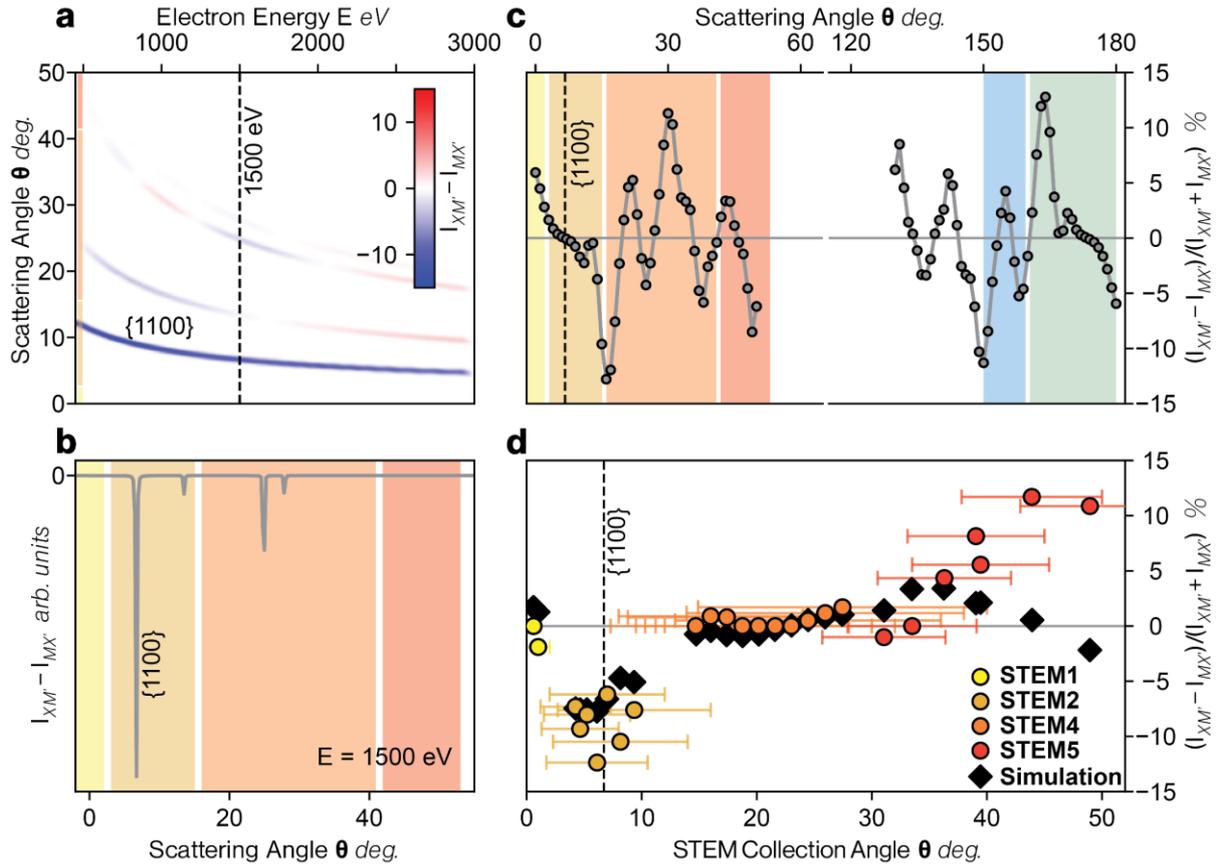

**Figure 2**. **Theoretical estimation of domain contrast in freely suspended samples and fitting to experimental data**. **a**) The difference in the elastic Bragg scattering transmitted intensity for the two domains, $I_{XM'}-I_{MX'}$ as a function of primary electron beam energy $E$ (eV) and the scattering angle, $\theta$. A colour scale is used to show the scattering intensity where white equates to no intensity. The dashed vertical line highlights E=1500 eV, the value used to the collect the experimental data points in (**d**). **b**) $I_{XM'}-I_{MX'}$ for E=1500 eV as a function of the scattering angle $\theta$. **c**) Domain contrast ($I_{XM'}-I_{MX'}$ / $I_{XM'}+I_{MX'}$) as a function of scattering angle for inelastically scattered electrons (note that angles greater than 90° represent reflected inelastically scattered electrons). Signal is radially averaged over both crystallographic directions for generated electrons with E=100 eV and a primary beam energy $E_0$ of 1500 eV. Colour shading in (**b**) and (**c**) matches the approximate angular ranges of the SEM detectors. **d**) Domain contrast from STEM detectors for a freely suspended twisted MoS$_2$ bilayer sample as a function of detector scattering angle. Each data point shows the mean domain contrast in a particular image. The horizontal bars indicate the annular range of scattering angles accepted by the detector for that image ($\Delta\theta$). The points are colour coded to match the STEM detectors used to acquire the image whose relative positions are shown in **Figure 1c**. The black data points are those calculated from theoretical modelling of the combination of elastic and inelastic scattering (using the same angular ranges as the equivalent experimental data point). The angle of the $\{\bar{1}100\}$ Bragg reflection at an incident electron energy of 1500 eV is indicated by the dashed vertical line.

*Domain contrast in the presence of a substrate*

While suspended bilayers can provide high contrast for domain imaging, it is often essential to investigate heterostructure samples supported on a substrate, where only reflected signals are available in the SEM. Bulk substrates are needed to provide mechanical support, back gating and for integration of heterostructures into manufacturing processes. To understand the effect of the substrate on image contrast, **Figure 3a** and **3b** compares simultaneously acquired ETD and EsB images where the field of view contains regions where the twisted bilayer is (i) fully suspended (inside the maroon circles), (ii) on an ultra-thin support (outside the maroon circle but inside the dashed green line) and (iii) on a bulk support (lower part of the images outside the green dashed line). The presence of an ultra-thin substrate (5 nm hBN) generates negative domain contrast in the ETD signal and enhances the positive domain contrast in the EsB signals compared to the full suspended region. Replacing the ultra-thin hBN with a bulk substrate (5nm hBN on top of 1um SiN) adds an additional background contribution to the EsB signal, slightly lowering the domain contrast relative to the ultra-thin case. However, surprisingly, we find the domain contrast of the ETD image inverts to show positive domain contrast for the bulk substrate (further detail in **SI Figure 2.4).** Consequently, the ultra-thin substrate gives inverted domain contrast for the ETD when compared to the EsB images, while the same contrast sign was always observed for all the EsB and ETD image pairs for twisted TMD layers on bulk supports. We note that similar bulk behaviour was observed irrespective of whether the support was 1 um SiN or 100 nm graphite on $SiO_2$/Si wafers (see **SI Figure S2.6**).

The contrast generated by thin and thick substrate conditions is shown schematically in **Figure 3c** and **3d**, respectively. To understand the domain contrast behaviour of supported samples, it is necessary to consider both (i) the interaction of the incident electron beam with the sample (as discussed in the first part of this work) and (ii) the interaction of the electrons that inelastically backscatter in the substrate and transmit through the bilayer in the opposite direction. Electrons backscattered from the substrate emerge through the sample and are detected by the EsB and ETD detectors. We can therefore reuse the learnings from the STEM detector modelling but with two major differences. Firstly, the backscattered electrons move through the bilayer in the opposite direction, so XM' and MX' domains now are inverted, producing opposite contrast sign in the elastic Bragg and inelastic contributions. Secondly, the backscattered electrons will have a wider spread of energies and angles than the primary beam. For elastic scattering, assuming an energy range of 500 – 1500 eV in the backscattered electrons, gives an angular distribution of the most significant $\{\bar{1}\,1\,0\,0\}$ Bragg reflection in the range of 168°< θ <173°, which is within the angular range of the EsB detector. These backscattered electrons that subsequently Bragg scatter as they pass back through the bilayer will therefore have positive domain contrast, adding to the positive EsB contrast from the suspended sample.

For a flat sample, the ETD collects electrons leaving the detector at a wide range of angles,[40] and therefore our calculations in **Figure 2c** show that both positive and negative contributions are possible depending on the angular range of the collected signal (which varies with the instrument settings and is difficult to quantify). We hypothesise that the angular range gives a negative contrast signal and that ultra-thin support increases the number of low energy electrons so that the negative domain contrast is visible above the detector noise in the ETD signal. For bulk supports the contrast inverts because this weak negative domain contrast signal is overwhelmed by the larger positive domain contrast due to the difference in the populations of the low energy electrons being generated as a secondary signal from the higher energy electron's positive contrast (illustrated schematically in **Figure 3d**).

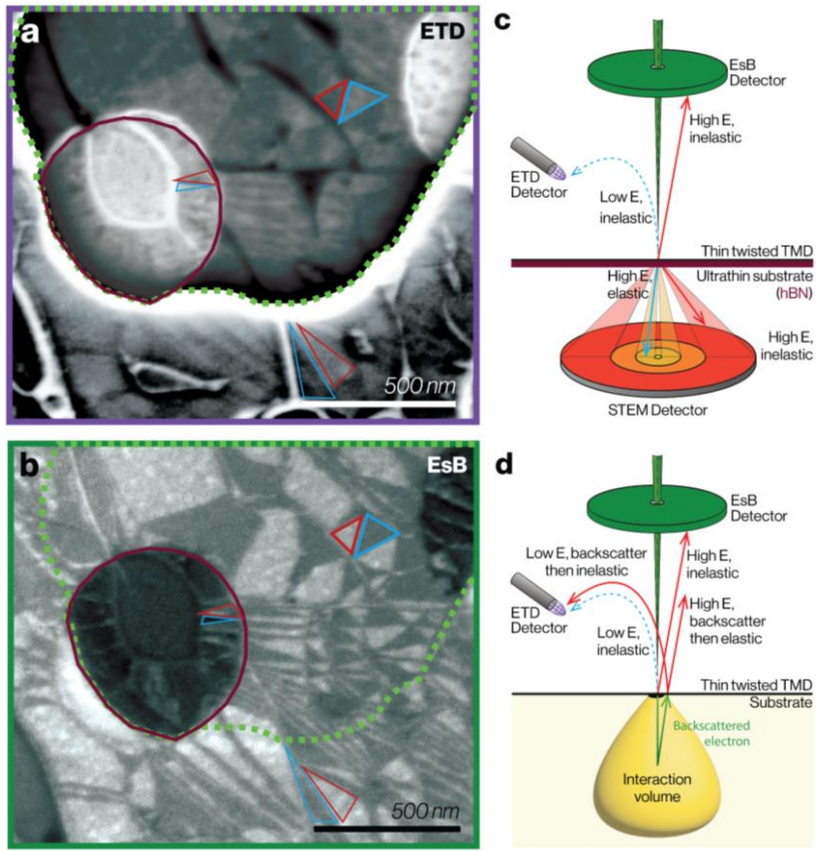

**Figure 3**: **Comparing domain contrast between images acquired with ETD and EsB electron detectors. a)** & **b)**, ETD and EsB images, respectively, for a twisted $WS_2$ bilayer sample which contains both substrate supported and suspended regions. The red region highlights a region where the bilayer is fully suspended and the area outlined by the green dashed line indicates where the bilayer is on an ultra-thin substrate (5nm hBN).The area outside the green dashed line the sample is on a μm-thick SiN support. In **a)** the ETD image is a montage of two images acquired with different dynamic range (optimised for the suspended region and fully supported regions, respectively, where the separate raw images are in **SI Figure S2.8**. **c)** and **d)**, Schematic depiction of detector geometry and measured contrast for (c) freely-suspended and (d) substrate-supported twisted TMD samples. Arrows denote experimental domain contrast, which is labelled as positive ($I_{XM} - I_{MX} > 0$, blue) or negative ($I_{XM} - I_{MX} < 0$, red).

## *Maximising contrast for twist domains*

One of the main instrument parameters for optimising SEM image contrast is the working distance (WD), which is the separation between the sample and the final probe forming lens.[41] Investigation of the effect of WD on domain contrast shows the behaviour that is expected for any SEM sample (**SI Figure S4.1a**, **b**). A WD of 7 mm provided the highest domain contrast for the ETD, while a slightly smaller value of 5 mm provided better contrast for the EsB image, with no contrast inversions observed for all the WD that were achievable. This suggests that the greatest domain contrast is achieved when the WD is set to maximise the number of collected electrons. A larger working distance improves the ETD detector collection efficiency but decreases the collection efficiency of the EsB and InL detectors. Smaller working distances favour higher spatial resolution as aberrations are reduced.[41] The optimal accelerating voltage for achieving the highest contrast for 3R twist domains in $MoS_2$ was found to be 1500 - 2000 V, with 1500 V used throughout this work for consistency. Further detail of optimising SEM imaging conditions is given in **SI section 4**.

*Domain contrast as a function of sample tilt*

In results presented to this point, we have demonstrated the significant domain contrast that is obtained for TMDs in ETD and EsB images even without specimen tilting. Yet previous work on channelling contrast imaging of 2D stacking domains relied on precise specimen tilting and azimuthal alignment with the ETD detector position to achieve contrast. [2,31] **Figure 4** shows that indeed tilting the sample has a strong effect on both the EsB and ETD domain contrast for twisted $MoS_2$ bilayers, and correct tilt conditions can be used to improve contrast. Channelling contrast is expected to vary with an azimuthal angle [2,30] (see calculations in **SI Figure S3.16**). We find that a tilt angle, τ, of ~21° gives the greatest domain contrast although compared to 0° there is a contrast is reversal at ~11° (**Figure 4b)**. To determine the optimal azimuthal direction for specimen tilt, the tilted sample was rotated until optimal contrast is achieved (see **SI Figure S4.8**). For BSE images it is only necessary to tilt along a specific crystallographic azimuthal direction, but for ETD imaging this crystallographic direction must be aligned towards the position of the ETD in the SEM chamber to prevent detector shadowing. **Supplementary Video 1** shows the domain contrast inversion as a function of sample tilt and **SI Figure S4.5** shows a similar effect of specimen tilt on domain contrast from $WS_2$. The observation that ETD and EsB signals oscillate in phase supports our conclusion that the main contributor to the ETD domain contrast is the low energy electrons produced by the higher energy channelling signal. The strongest domain contrast observed here was from the EsB detector, which for an MoS$_2$ twisted bilayer reaches -11% at the optimal tilt of 21°.

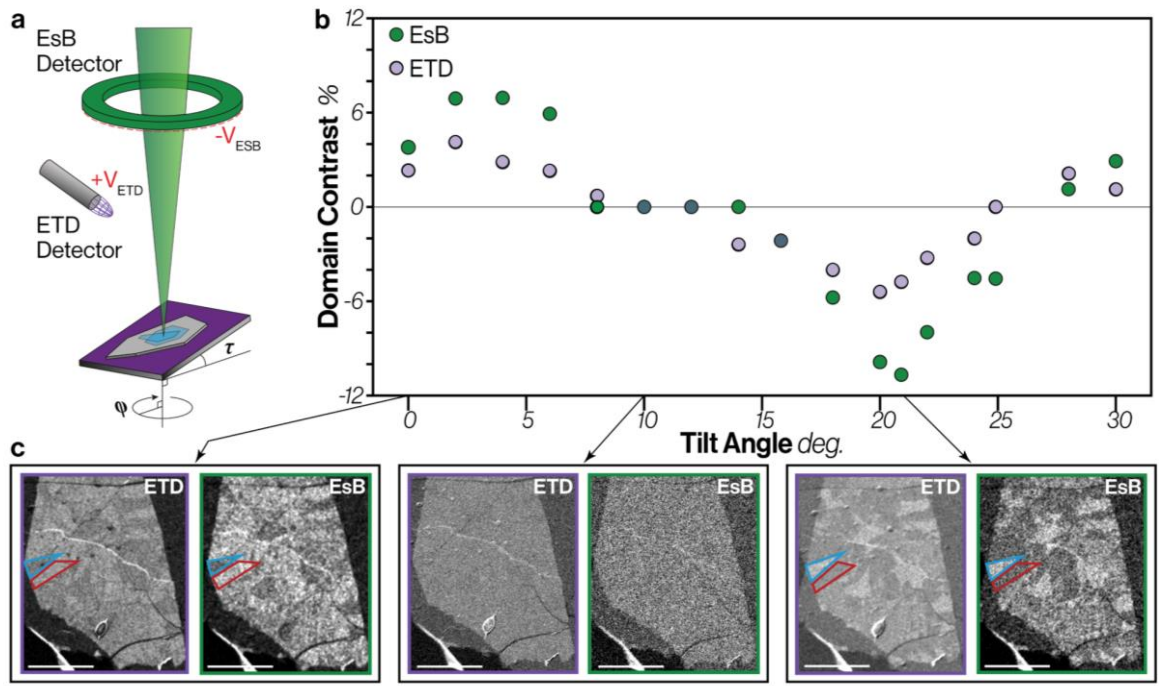

**Figure 4: Domain contrast as a function of tilt angle** for a 3R type MoS$_2$ bilayer on graphite and an oxidised silicon wafer. **a**) Schematic diagram with tilt (τ) and rotation (Φ) angles shown. **b**) Comparison of domain contrast as a function of specimen tilt angle for the ETD (SE) and EsB (BSE) detectors. (c) the ETD and EsB image pairs for stage tilt angles of 0° (positive channelling contrast), 10° (no contrast) and 22° (negative channelling contrast). Tilted images have been corrected for tilt distortion. All scale bars are 3 μm.

*Twist domain contrast with encapsulation*

We now consider the applicability of the BSE ECCI approach for imaging stacking domains in realistic electronic device architectures which require encapsulation, capping layers or top-gates. Previous work has shown that ETD images can show stacking differences in twisted WSe$_2$ bilayer encapsulated in 5 nm of graphene/hBN, but it was necessary to tilt to 40° and use an electron beam energy of 3000 eV.[31] Furthermore, no quantification of the domain contrast degradation due to encapsulation was presented. Here we quantitatively compare the change in domain contrast after encapsulation by fabricating a twisted MoS$_2$ bilayer with a 3.5 nm thick (10 layer) hBN encapsulation layer that only covers part of the sample: the edge is indicated by the dashed lines in **Figure 5a**, **b**. For both the ETD and EsB images at zero tilt, hBN encapsulation decreases the positive contrast by approximately 50% compared to the unencapsulated case (**Figure 5a**). At ~21° tilt, the negative domain contrast is reduced by approximately 65% for both signals (**Figure 5b**), where the further reduction relative to that seen at zero tilt can be explained by the increased thickness of the hBN seen when the sample is tilted. The observation that both ETD and EsB signals are reduced by the same amount again supports our conclusion that the ETD signal originates from channelled higher energy electrons; if the signal originated from channelling of low energy electrons they would not escape from the encapsulation layer (since SE with energies of less than 50 eV have an escape depth less than 1 layer of $MoS_2$ or less than 1 nm hBN, see **SI Figure S3.9**).[39] These results provide further evidence that the domain contrast in the ETD signal is generated by the channelling of higher energy electrons backscattered through the bilayer.

Surface contamination will also act to degrade the domain contrast compared to perfectly clean surfaces. The extent of this contamination induced degradation, resulting from polymer transfer layers or electron beam deposited hydrocarbon layers, can be expected to be similar to that quantified for hBN for equivalent thicknesses, as both types of contamination have similar mean atomic number to hBN. We have found that *in situ* plasma cleaning is able to reduce the deposition of surface contamination during serial SEM imaging and can even facilitate the cleaning of a contaminated surface by the electron beam resulting in improved contrast for twist domains (see **SI Figure S4.9** and **Supplementary Video S3**).

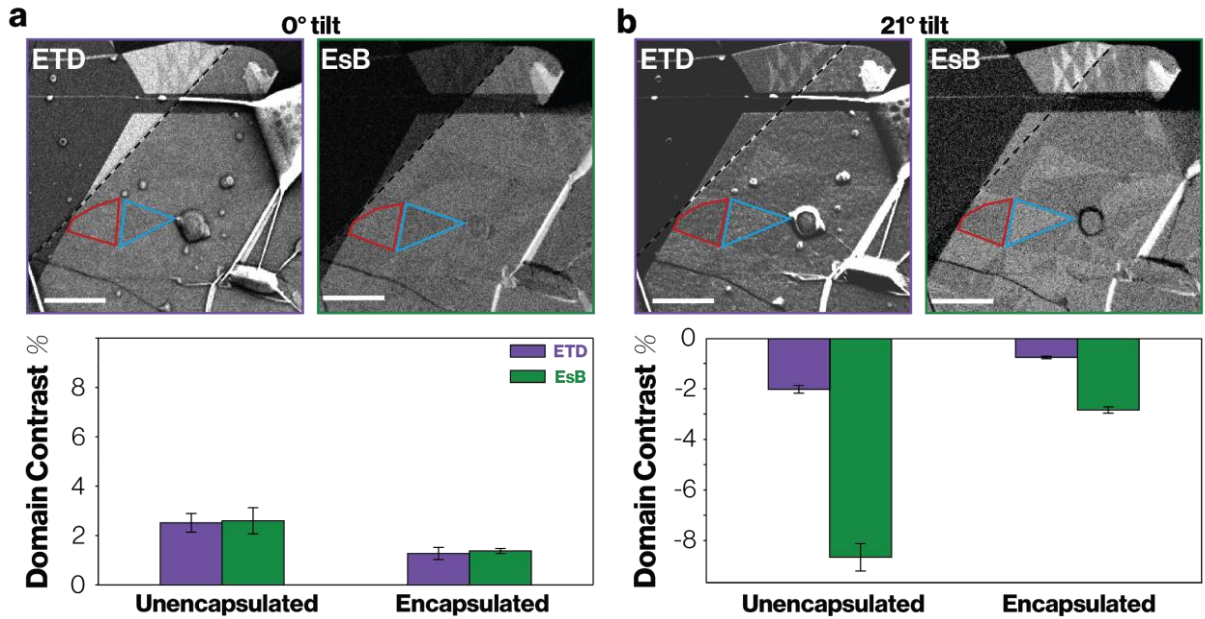

**Figure 5: Change in domain contrast with and without 10-layer hBN encapsulation** at **a**) 0° and **b**) 21° specimen tilt, respectively. The top row consists of ETD and EsB images of moiré superlattice domains in twisted bilayer MoS$_2$ where a 3.5nm hBN encapsulation layer covers only the lower part of the specimen. The edge of the hBN flake is indicated by the black/white dashed line, hBN thickness of 3.5 nm was determined *via* atomic force microscopy. Scale bars are 1µm. Bottom row shows the quantified mean contrast between domains (note the inversion of domains between 0 and 21° specimen tilt consistent with the data in Figure 4).

*Twist domain contrast for thicker samples and other TMDs*

**Figure 6** compares the effect on twist domain contrast when the lower TMD layer is thicker (a monolayer on a pristine bilayer or monolayer on a pristine trilayer). These heterostructures have reduced domain contrast compared to the bilayer consisting of a monolayer on a monolayer. We assign this to the increased thickness of the lower layer adding to the background signal without adding to the channelling difference between the domains (**Figure 6c**). Nonetheless, we find that the EsB domain contrast signal for twist domains in WS$_2$ remains high (above 20%), even when the lower WS$_2$ crystal is 3 layers thick. For this sample/imaging configuration the EsB signal gives greater domain contrast than the ETD signal for all heterostructures, but the additional thickness of the lower layers decreases the observed contrast of both detectors by a similar factor, further supporting our hypothesis that the ETD signal is coupled to the relative populations of higher energy electrons emerging from the sample.

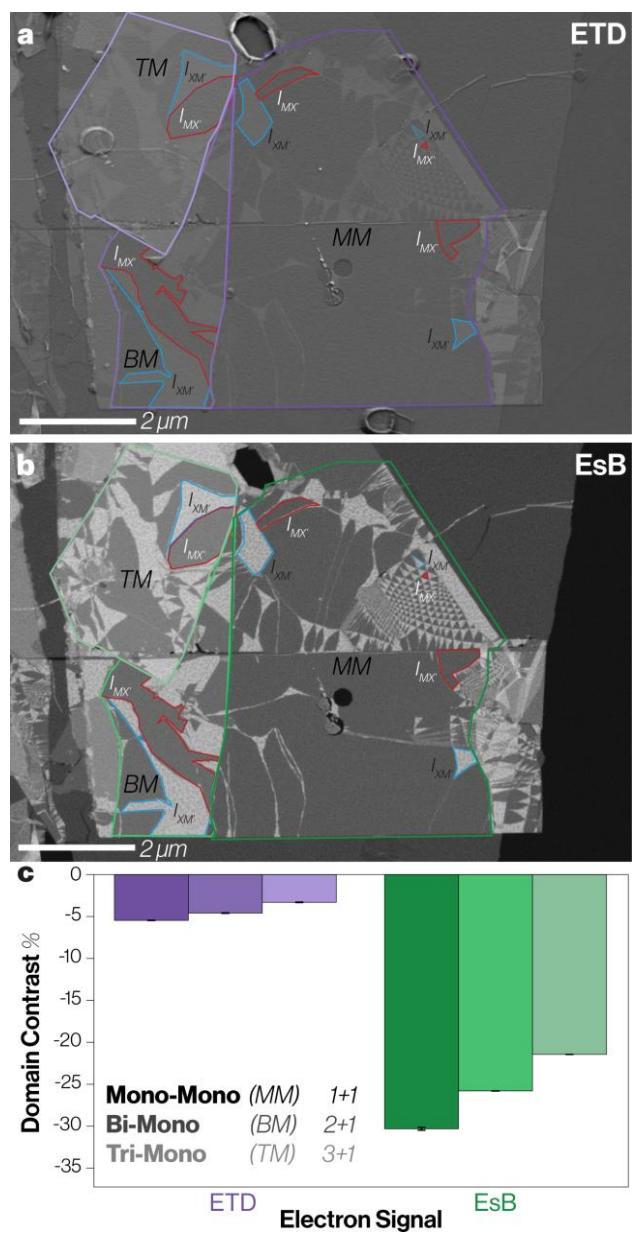

**Figure 6**: **Domain contrast from a) ETD and b) EsB images for twisted WS$_2$ heterostructure regions including TMD layers with different thicknesses** (sample is unencapsulated). Different thicknesses of the lower TMD layer are indicated by the outlines in **a** and **b**. The light-to-dark outlines highlight monolayer-trilayer (TM), monolayer-bilayer (BM), and monolayer-monolayer (MM) domain thicknesses, respectively. For each region, the specific domains used for measuring $I_{XM'}$ and $I_{MX'}$ values are shown *via* the red and blue polygons, respectively. **c)** The twist domain contrast as a function of bottom TMD layer thickness. Imaging was performed with a working distance of 5.5 mm, an acceleration voltage of 1.5 kV, and a stage tilt of 20.2°.

**Conclusions**

This work demonstrates the largely unexploited potential of conventional SEM imaging for revealing reconstructed twist domain networks in TMD heterostructures. We find that channelling contrast SEM imaging of twisted domains is achievable for most SEM models and configurations (examples shown in **SI Figure S4.10**), without requiring prior knowledge of sample crystallinity, supporting the widespread opportunities for applying the ECCI approach for imaging stacking domains in TMDs.

Our calculations demonstrate that for transmitted electrons in suspended bilayers, domain contrast inversion occurs at moderate scattering angles, with positive contrast from inelastic electrons dominating at high angles and negative contrast from elastic interactions (mainly the {1-100} Bragg reflection) important at low angles. Optimal domain contrast for transmitted electrons at 1500 V accelerating voltage and zero specimen tilt is thus achieved by choosing an annular STEM detector with either (i) a large collection angle and an inner detector angle of ~25° to exploit the positive contrast from inelastically scattered electrons or (ii) a detector that has a small angular range to exclude the directly transmitted beam and the electrons from high-angle scattering and a collection angle centred on the $\{\bar{1}\,1\,0\,0\}$ Bragg reflection to achieve negative contrast. The EsB detector also gives strong positive domain contrast by collecting the reflected inelastic BSEs from suspended samples.

Both $MoS_2$ and $WS_2$ bilayers on conventional graphite/$SiO_2$ substrates give SEM images with positive twist domain contrast even without specimen tilt (when imaging along the [0001] direction) for all detectors (EsB and InL and ETD). Our calculations suggest that the positive domain contrast seen on the EsB detector for the suspended sample is enhanced by the presence of the substrate, which adds positive contrast contributions from both elastic and inelastic scattering contributions. The situation is more complicated for the ETD signal where the weak negative contrast observed for suspended samples is overwhelmed and ETD detector signal for supported samples is due to low energy secondary electrons generated by the higher energy electron signals. Thus, the domain contrast of TMDs on bulk substrates always behaves in a similar way whether high or low energy electrons are being detected.

A specimen tilt of 21° away from the [0001] direction can be applied to increase domain contrast by ~4 times relative to the 0° case (accompanied by an inversion in domain contrast).

WS$_2$ bilayers have approximately double the contrast of MoS$_2$ resulting in twist domain contrasts of -31% and -16% for WS$_2$ and MoS$_2$, respectively, at the optimal tilt angle of 21°. Importantly for practical use to image electronic structures, the twist domains are visible for both WS$_2$ and MoS$_2$ bilayers beneath a 3.5nm hBN encapsulation layer, in the presence of hydrocarbon surface contamination, and for complex heterostructures including trilayer TMDs. Although imaging twist domains with the conventional ETD SE imaging mode is entirely possible, we find the EsB, InL and other annular BSE detectors are generally preferrable for several reasons. Firstly, annular BSE detectors generally allow smaller working distances and therefore higher spatial resolution to be achieved than is possible for ETD detectors. Secondly, the optimal sample tilt does not need to be towards the ETD detector. Thirdly, the higher energy electron signal is less sensitive to surface topography and charging so is also capable of revealing domain contrast where the low energy signal fails, such as where there are encapsulation bubbles or for insulating heterostructures.

We believe this readily accessible and inexpensive approach to image twisted domain configurations in fabricated (opto)electronic devices has great potential to provide new understanding of variations in device performance and for optimising device configuration, as well as for *in situ* SEM measurements of moiré superlattices and variable stacking configurations in TMDs.

## Methods

**Sample preparation:**

Samples were prepared from mechanically exfoliated crystals *via* a modified tear-and-stack method.[34] Specific details can be found in **SI section 1**. The use of metal-coated substrates or lithographic electronic contacts is recommended as charging becomes a significant obstacle to the visualisation of twisted 2D materials with SEM, as illustrated in **SI Fig 2.7.**

**Scanning electron microscopy:**

The primary SEM used for this investigation was the Zeiss Merlin Gemini II. An acceleration voltage of 1500 V, beam current of 1 nA, a pixel dwell time of 52 µs, working distance of 4-5 mm, and a 400V and -800V detector bias for the ETD and EsB detectors, respectively, were used when modulating channelling contrast with stage parameters. Detector annular collection ranges are 179°-160° degrees for the EsB, 160°-150° for the InL (angles measured from the primary beam forward direction) for a primary beam energy of 1500 eV. The STEM detector angular range varies with detector-sample distance, in the range of 1.2°-16.0° for STEM2 and 25.6°-55.0° for STEM5 (angles measured from the directly transmitted beam for a detector-sample distances of 5.8-13.0 mm and 5.7-11 mm, respectively). The scanning transmission electron microscopy (STEM) data collected with the Merlin's segmented STEM detector. Other microscopes used include an FEI Magellan, Apreo SEM, Helios 660 FIB SEM and a Zeiss Ultra with similar microscope parameters as listed above.

**Atomic force microscopy (AFM):**

Amplitude-modulated (AM) tapping mode AFM was performed with an Asylum Research Cypher-S AFM in ambient conditions. Images were acquired using budget sensors ElectriMulti 75-G probes with a nominal force constant of 3 N/m and resonant frequency of 75 kHz. The imaging was performed with an amplitude setpoint of 60-80% of the free-air amplitude and driven at -5% of the resonant frequency of the cantilever with an average force setpoint of 10-40 nN. All AFM data analysis was performed using the open-source Gwyddion software package.

## Acknowledgements

S.J.H. and E.T. thank the Engineering and Physical Sciences Research Council (EPSRC) for funding under grants EP/S021531/1, EP/M010619/1, EP/V001914/1 and EP/P009050/1, the National Physical Laboratory (NPL) UK and the European Research Council (ERC) under the European Union's Horizon 2020 research and innovation programme (Grant ERC-2016-STG-EvoluTEM-715502). TEM access was supported by the Henry Royce Institute for Advanced Materials, funded through EPSRC grants EP/R00661X/1, EP/S019367/1, EP/P025021/1 and EP/P025498/1.

**Supplementary Information**

**Electron Channelling Contrast SEM Imaging of Twist Domains in Transition Metal Dichalcogenide Heterostructures**

*Evan Tillotson[1,3], Dr. James McHugh[2,3], Dr. James Howarth[3], Dr. Teruo Hashimoto[1], Dr. Nick Clark[2,3], Dr. Astrid Weston[2,3], Dr. Vladimir Enaldiev[2,3], Samuel Sullivan-Allsop[1,3], William Thornley[1,3], Dr. Wendong Wang[2,3], Dr. Matthew Lindley[1], Dr. Andrew Pollard[4], Prof. Vladimir Falko[2,3^], Prof. Roman Gorbachev[2,3#] and Prof. Sarah J. Haigh[1,3*].*

1. Department of Materials, University of Manchester, Manchester M13 9PL, UK
2. Department of Physics and Astronomy, University of Manchester, Manchester M13 9PL, UK
3. National Graphene Institute, University of Manchester, Manchester M13 9PL, UK
4. National Physical Laboratory, Hampton Rd, Teddington, TW11 0LW, UK

sarah.haigh@manchester.ac.uk [*]; roman@manchester.ac.uk [#]; falko@manchester.ac.uk [^].

# Contents

# 1. Sample Preparation

A modified version of the tear-and-stack technique[1] was used to prepare the 3R-type twisted TMD samples on a Si/SiO$_2$ wafer substrate, using a micromanipulation transfer rig in an argon atmosphere glove box. The monolayer and few-layer TMD flakes were prepared by mechanically exfoliating bulk crystals onto Si/SiO$_2$ wafers spin coated with polypropylene carbonate (PPC). The micromanipulator needle was employed to tear the exfoliated flake in half, so that a poly-methyl methacrylate (PMMA) carrier layer could be deposited to selectively pick up one half of a flake (at 55 °C) and deposit it on to the second half with a specified rotation. The 3R-type homo-bilayers were then either deposited onto a graphite crystal on a Pt (50 nm)/Ti (5 nm) coated Si/SiO$_2$ wafer, or on to a thick hBN crystal. To achieve hBN-encapsulation, a further hBN crystal (2-3.5 nm thick as determined by AFM) was then transferred using a PMMA carrier layer. In the absence of the metallic Pt-Ti layer, the bilayer was grounded *via* lithographically patterned electrodes. The whole 2D van der Waals heterostructure was then annealed in a vacuum chamber (pressure approximately 10$^{-6}$ mbar) at 200°C to minimise surface contamination prior to SEM imaging. To produce the freely suspended heterostructure samples the twisted bilayers were picked up using a patterned hBN crystal and transferred to a holey SiN$_x$ support.

# 2. Quantification of Domain Contrast

Here we employ Michelson contrast, $C_m$ to compare domain contrast for different conditions:

$$C_m = 100 \frac{I_d}{I_s} = 100 \frac{I_2 - I_1}{I_2 + I_1}$$

Where $I_1$ and $I_2$ are the mean intensity of the two stacking domains, $I_d$ is the intensity difference ($I_1 - I_2$) and $I_s$ is the intensity sum ($I_1 + I_2$). The mean intensity value for a specific domain region, $I_1$ or $I_2$, was determined by extracting pixel by pixel intensity histograms for a single domain using the ImageJ software. These histograms also allowed measurement of the standard deviation in the individual pixel intensities for the individual domain ($\Delta I_1$ and $\Delta I_2$). To determine the measurement errors for the reported contrast values $\Delta C_m$, standard deviation errors have been propagated in accordance with standard theory:

$$\Delta C_m = \sqrt{\left(\frac{\partial c_m}{\partial I_1}\Delta I_1\right)^2 + \left(\frac{\partial c_m}{\partial I_2}\Delta I_2\right)^2} \quad \therefore \quad \Delta C_m = \sqrt{\left(\frac{200\Delta I_1 I_2}{(I_1+I_2)^2}\right)^2 + \left(\frac{200 I_1 \Delta I_2}{(I_1+I_2)^2}\right)^2}$$

**Figure S2.1** provides an overview of the method used to measure individual domain contrast for an MoS$_2$ bilayer region (A1). First the domain region in the image is isolated. Any contamination bubbles or defects present (see **Figure 1a**) are excluded from the region of interest (see **Figure S2.1c**) so they do not contribute to the intensity histogram. Selecting the pixels that contribute to a single domain enables measurement of the intensity histogram whose mean intensity value and standard deviation provide the necessary values for $I_1$ and $\Delta I_1$, respectively. The process is repeated for a second domain with the different contrast as near as possible to the first domain to measure $I_2$ and $\Delta I_2$, enabling $C_m$ and $\Delta C_m$ to be determined.

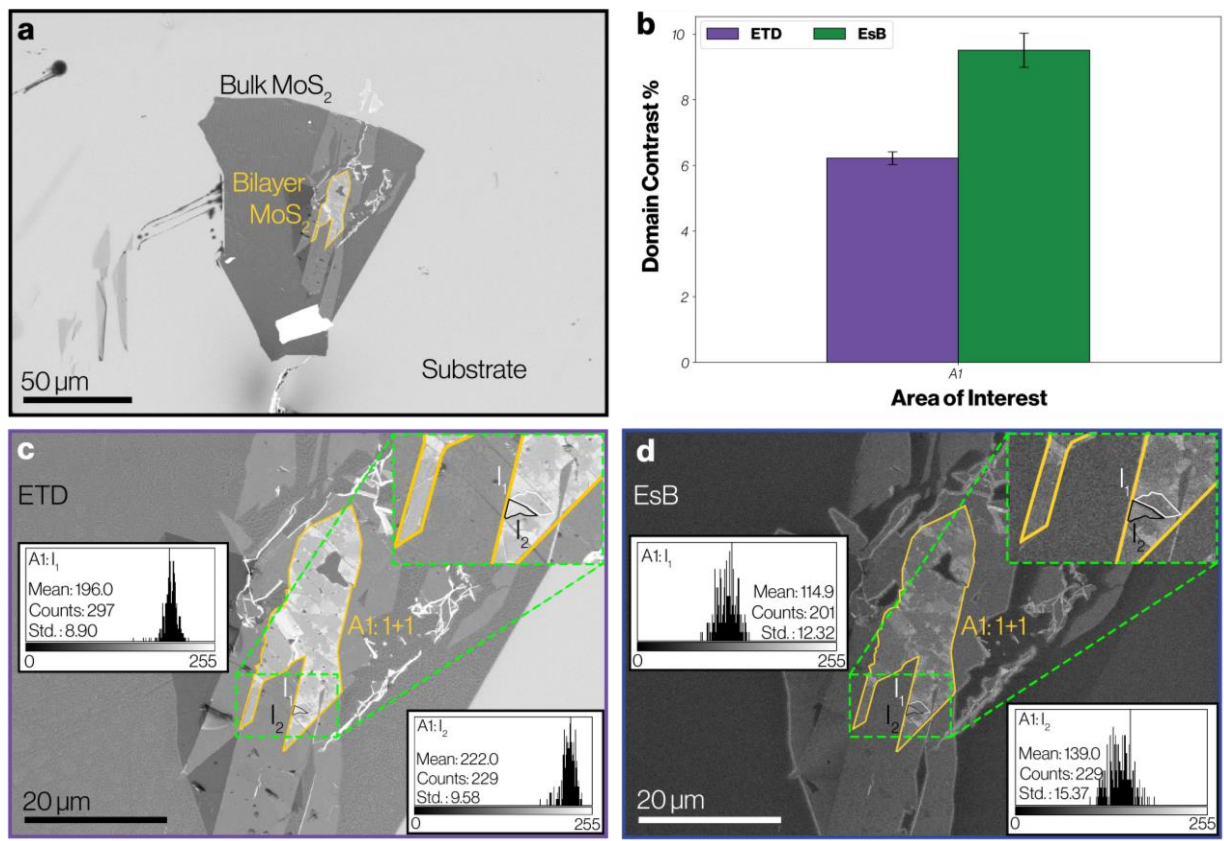

**Figure S2.1: Method of measuring the Michelson contrast of reconstructed twist domains in an MoS$_2$ twisted bilayer (not encapsulated). a**) An ETD SEM overview image with highlighting the bilayer of MoS$_s$ (outlined in yellow) within the bulk flake. **b**) Plot of the Michelson contrast for the twisted domains measured from ETD and EsB images in area A1. Imaging parameters: 5.9 mm working distance, 1.0 kV acceleration voltage, 0.1° stage tilt. **c**) & **d**) ETD and EsB images of the region containing the twist domains highlighted by the yellow boundary in panel **a**). The insets (green border) in the top right show a magnified region where the individual domains are outlined in white and black, having intensities of $I_1$ and $I_2$ respectively. Also inset are intensity histograms from the individual domains ($I_1$ and $I_2$) with the mean and standard deviation (std.) values, along with the number of counts (pixels) shown. Often the separate domain intensities were not distinguishable in the intensity histogram of the full area (yellow polygon) so the evaluation of bright/dark domain intensities and their standard deviations was performed by measurements made from individual domains using ImageJ.[2]

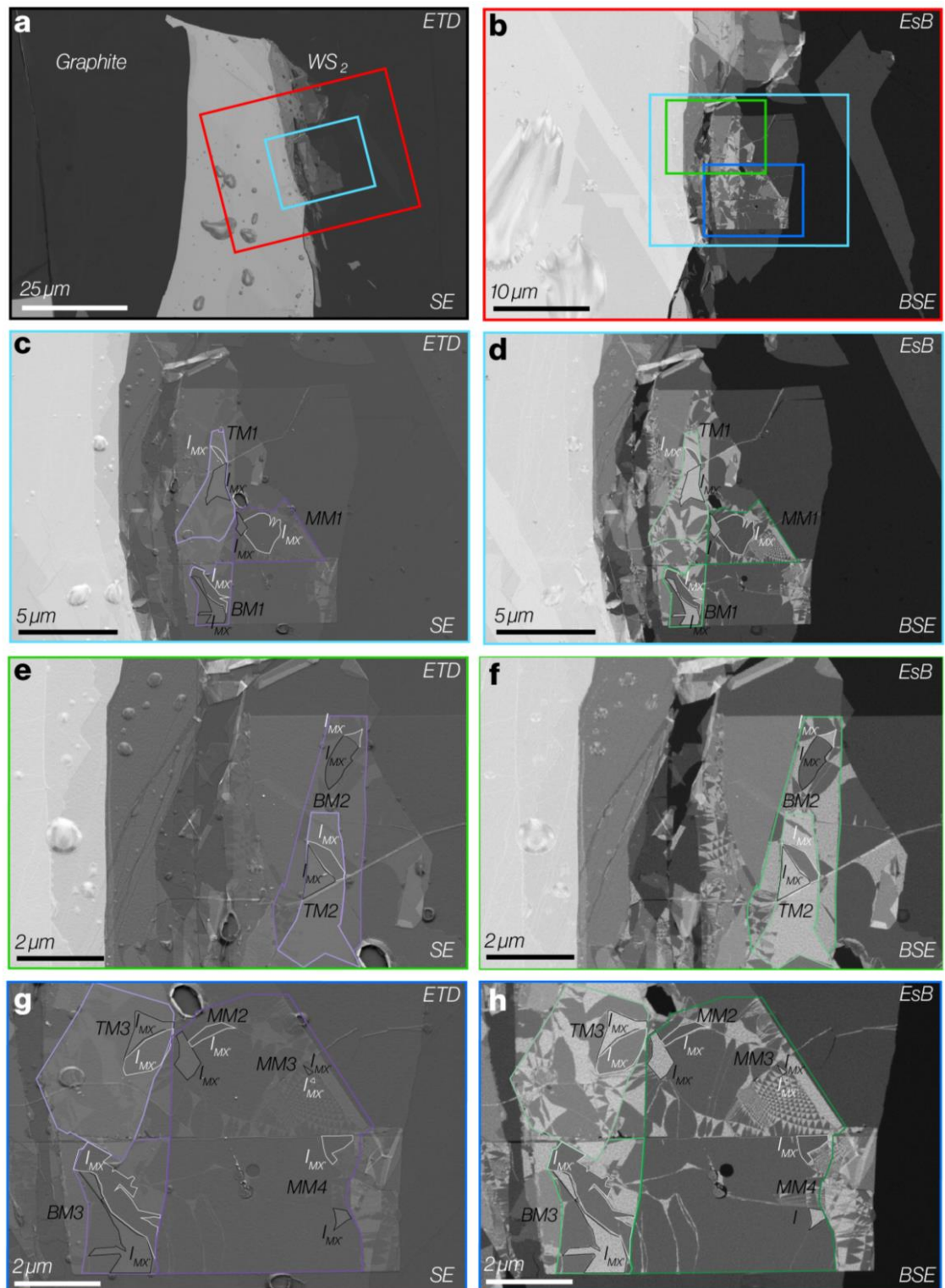

**Figure S2.2: Comparison of domain contrast from a twisted WS$_2$ heterostructure including different thicknesses for the bottom layer (not encapsulated).** Left column, **a**, **c**, **e** and **g**, are ETD images. Right column, **b**, **d**, **f** and **h**, are EsB images. Different thickness regions are indicated *via* differing shades of coloured outlines in purple and green for ETD and EsB images, respectively. The lightest shading is monolayer on trilayer (TM). The medium shading corresponds with monolayer on bilayer (BM) regions while the darkest shading highlights monolayer on monolayer (MM) areas. For each region, the specific domains used for measuring I$_1$ and I$_2$ values are shown *via* the black and white polygons, respectively. The intensity histograms of these domains can be found in **Figure S2.3**. Imaging parameters 5.5 mm working distance, 1.5 kV acceleration voltage, 20.2° stage tilt.

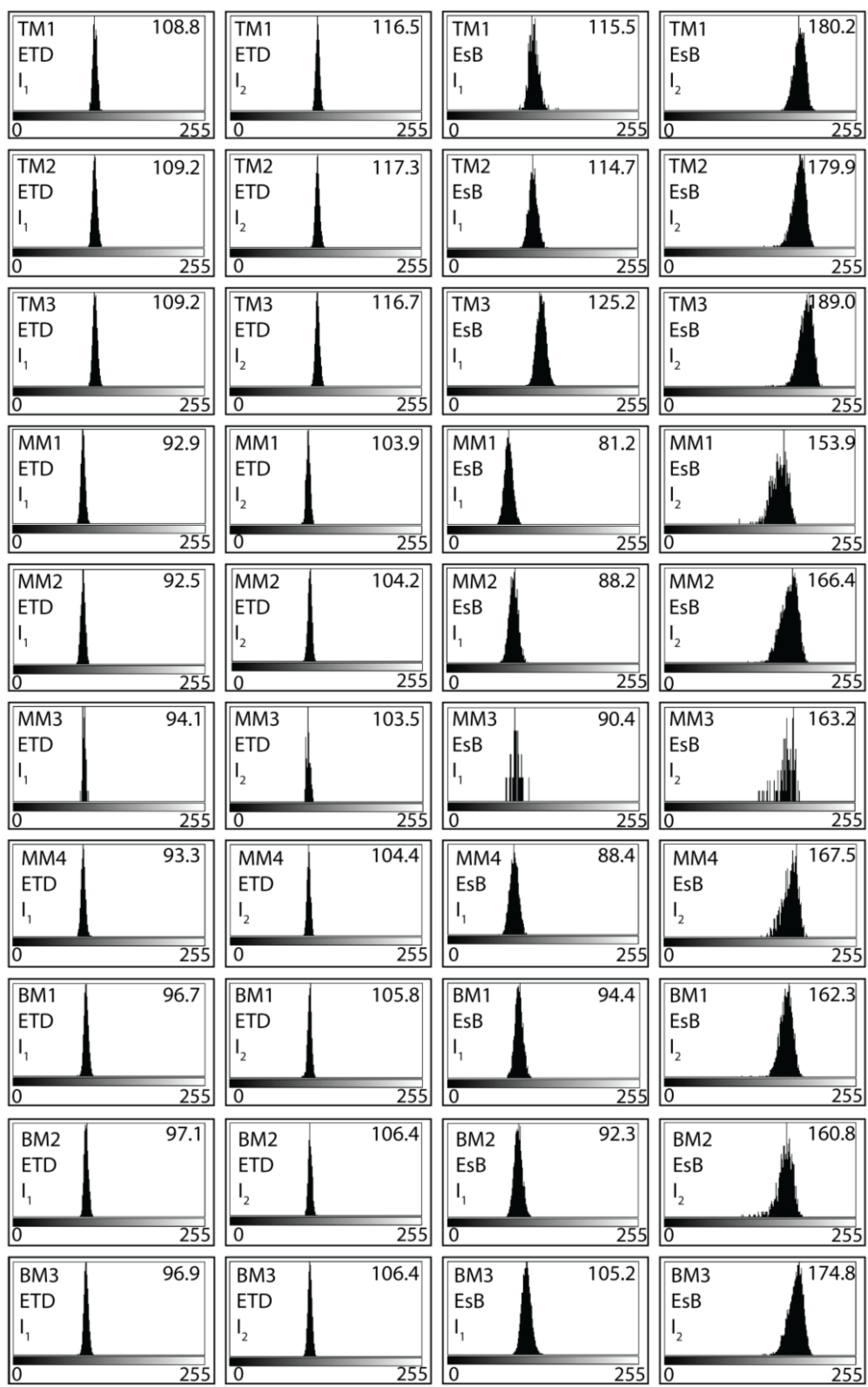

**Figure S2.3: Intensity histograms measured from the WS$_2$ domains outlined in Figure S2.2.** All images were analysed as 8-bit TIFs (0-255). Mean intensity values are given top right for each histogram.

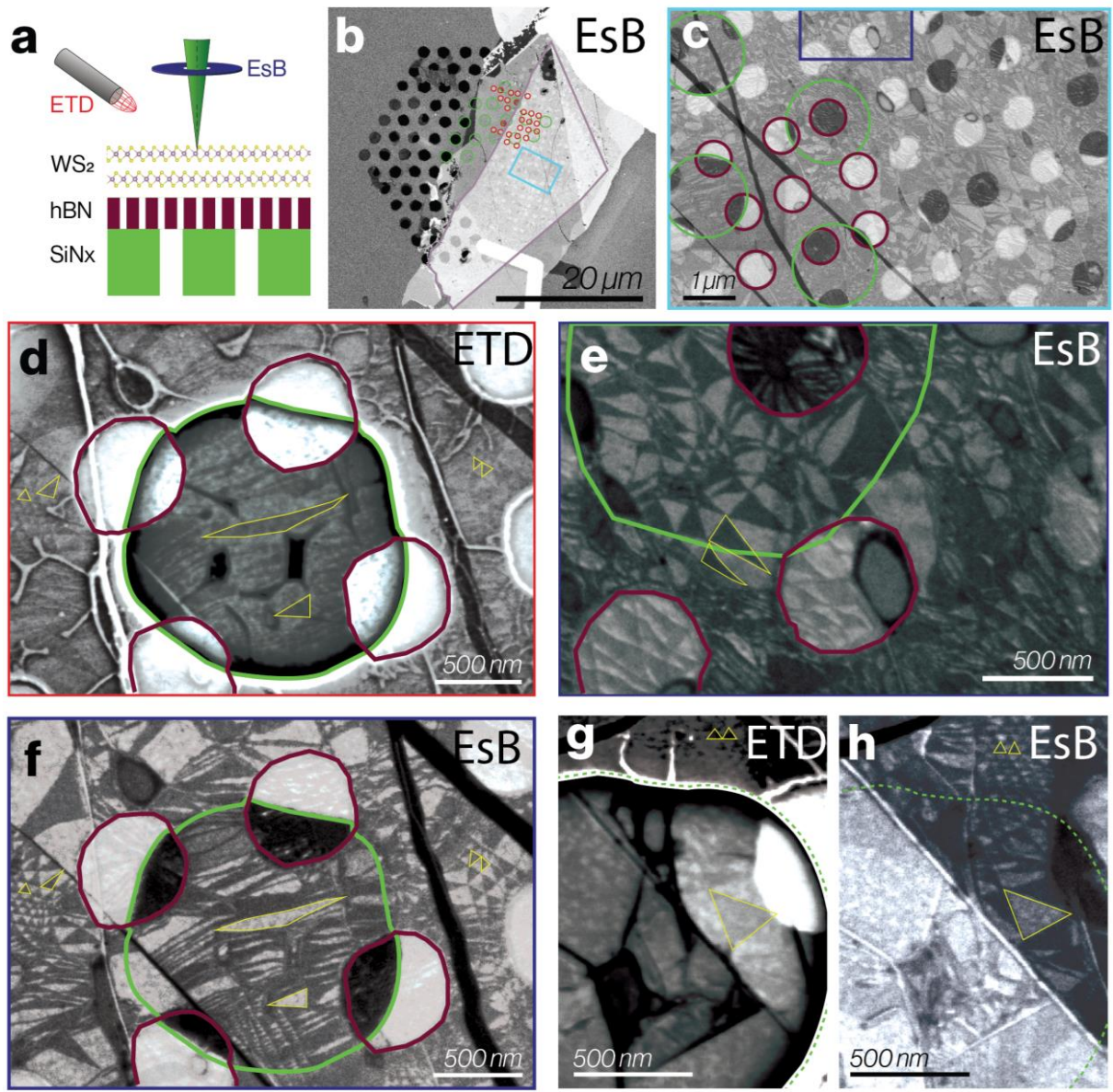

**Figure S2.4: Comparison of ETD (red) and EsB (blue) electron detector signals for a hBN supported WS$_2$ bilayer. a**) Schematic cross-section of the suspended WS$_2$ bilayer supported on a hBN substrate with 500 nm diameter holes (red borders) and a SiN$_x$ TEM grid with 2μm diameter holes (green borders). **b**) and **c**), EsB images of double-suspended twisted WS$_2$ bilayer at low and high magnifications, respectively, where **c**) corresponds with the blue rectangle highlighted in **b**). **d**) & **g**) ETD images and **f**) & **h**) EsB images showing contrast inversion within pseudo-freely-suspended regions when comparing SE and BSE images. **e**) BSE image showing retention of contrast from freely-suspended to substrate-supported areas.

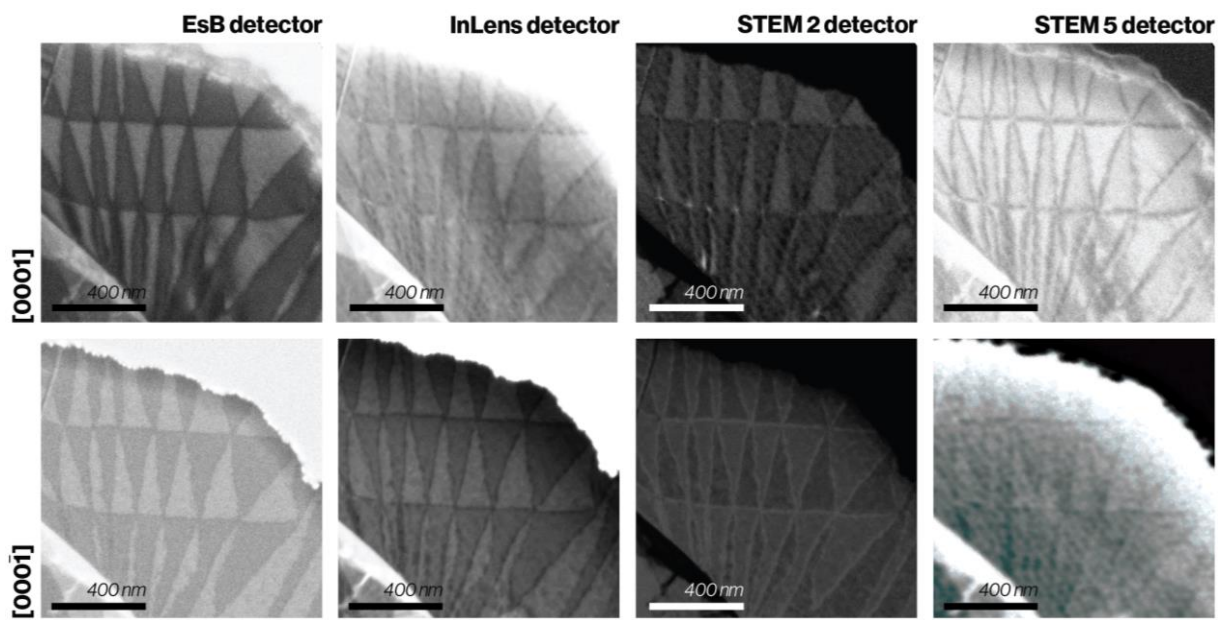

**Figure S2.5: Contrast inversion achieved by turning the sample upside down. Top row: SiNx holey support side down. Bottom row: SiN$_x$ holey support side up.** Identical samples are viewed along close to either the [0001] (top) or [000$\bar{1}$] (bottom) crystallographic direction. Identical detector conditions are used for both orientations. Contrast for STEM5 detector is reduced in the [000$\bar{1}$] images, likely due to surface contamination.

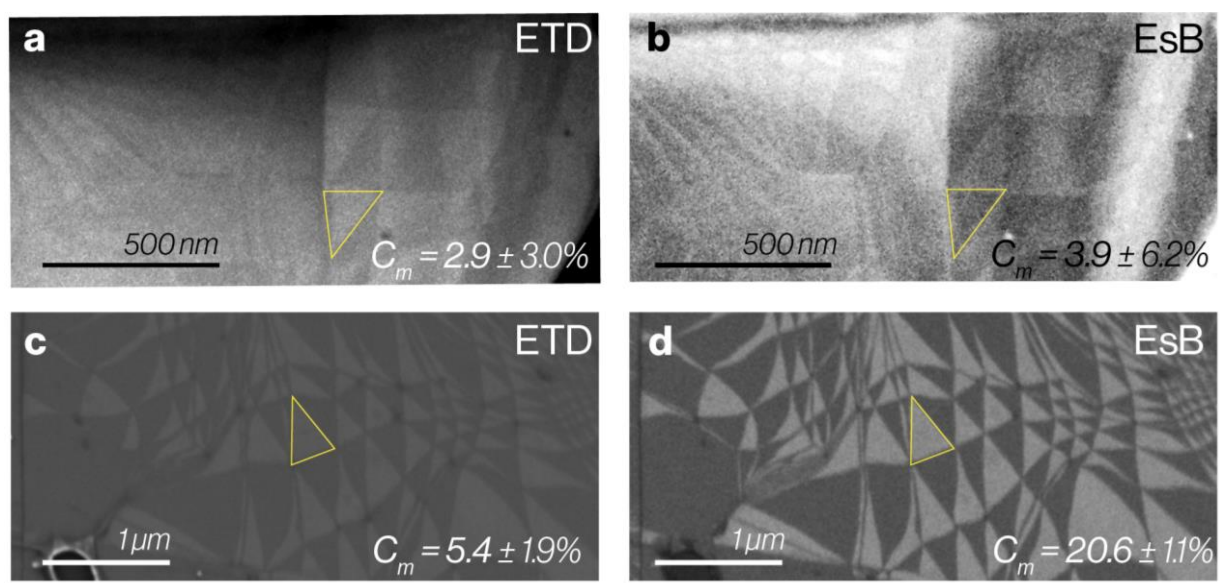

**Figure S2.6: Comparison of ETD and EsB signals for suspended and supported WS$_2$ twisted bilayers. a**) & **b**) are suspended samples, indicating contrast inversion between the two imaging modes. **c**) & **d**) are twisted bilayers on a solid graphite/SiO$_2$ substrate the two signals have qualitatively similar domain contrast.

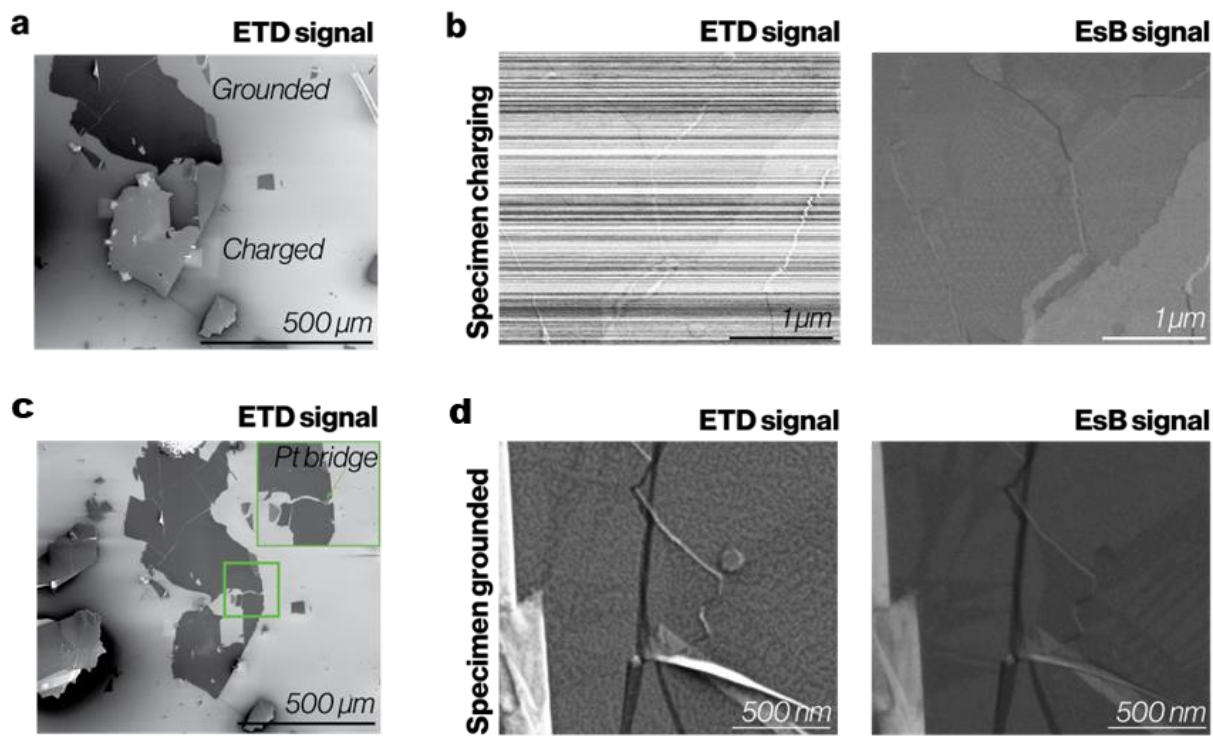

**Figure S2.7: Effect of specimen charging on ETD and BSE images. a**) Shows the sample as initially fabricated, where the flake towards the bottom of the image is charging (different contrast observed between the flakes. **b**) Higher magnification imaging of this charging flake is possible with the EsB detector but not with the ETD. **c**) Shows the same region as in a) after the bottom flake has been grounded by an electron beam deposited platinum strip. The inset shows an enlarged view of the Pt bridge. **d**) After the Pt strip deposition both ETD and EsB signals are restored. However, the topography caused by unintended Pt deposition outside the region of the Pt strap obscures the visibility of the channelling contrast in the ETD signal.

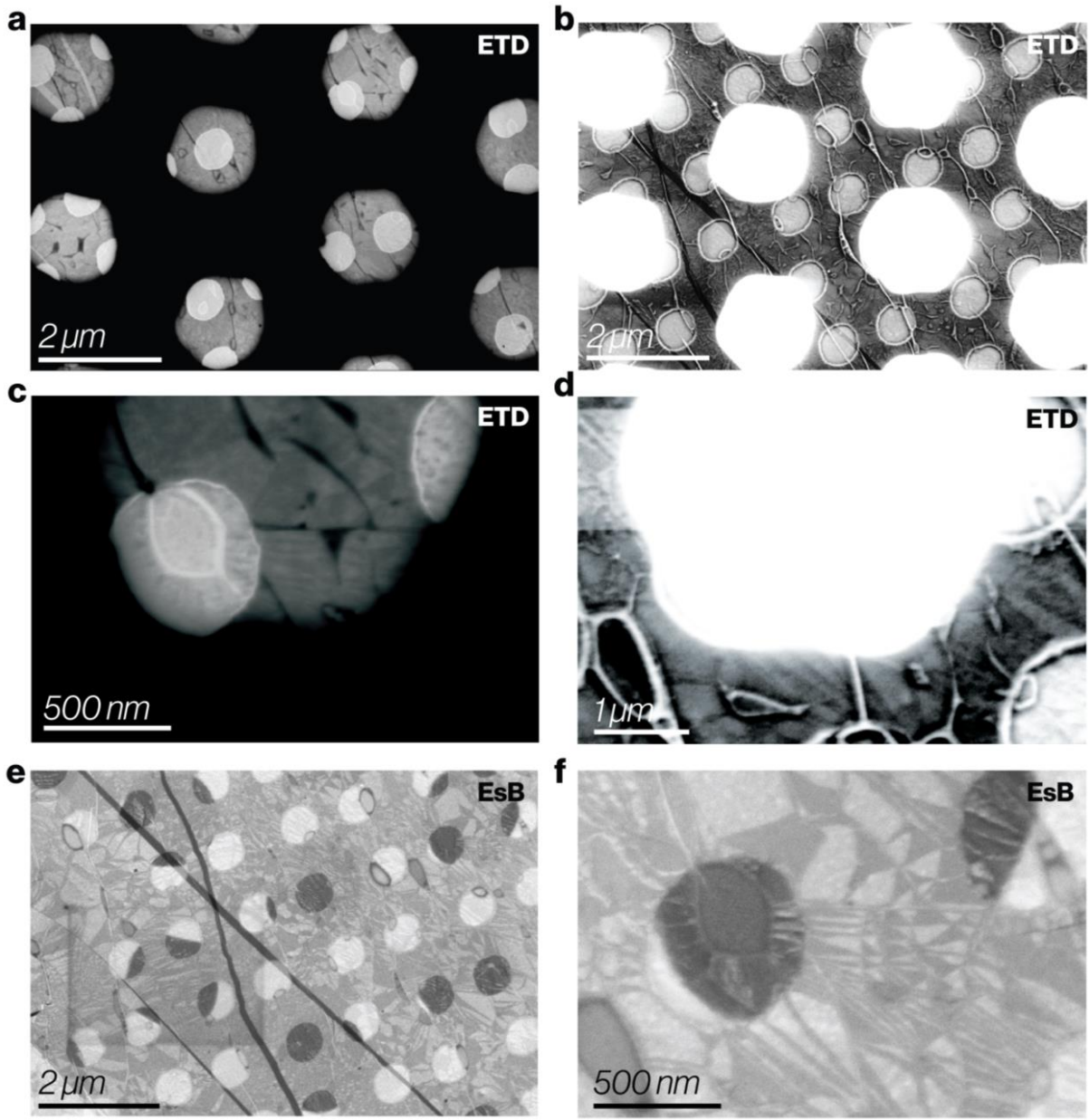

**Figure S2.8: Example images showing the twisted TMD bilayer with fully suspended areas, ultra-thin support areas and areas on a bulk support within the same field of view. Small circles are the holes in the ultra-thin hBN membrane support. Larger holes are in the bulk SiNx support membrane. The ETD images in (c & d) are the raw images used to produce the composite image with higher dynamic range shown in Figure 3a (main text). a,b**) and **c,d**) are pairs of ETD images showing the same area but with different gain applied to the ETD detector. **e**) & **f**), Are the EsB images corresponding with **a,b**) and **c,d**), respectively.

# 3. Modelling of Electron Scattering as a Function of Collection Angle

### 3.1. Elastic Scattering
The total electron scattering contains both elastic and inelastic scattering processes which need to be considered differently (see **Figure S3.1**).

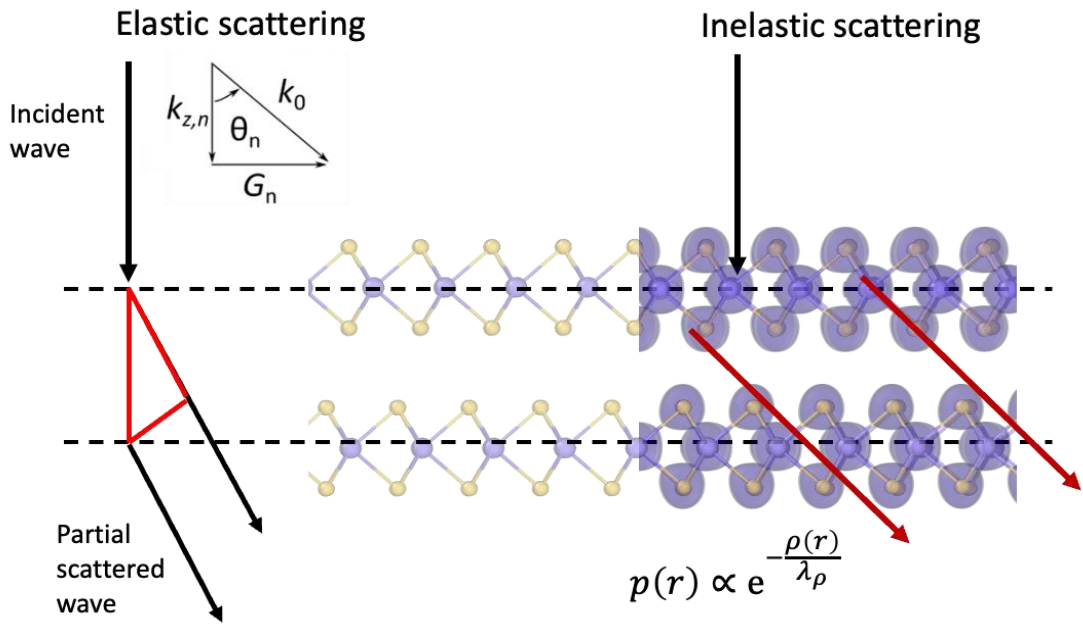

**Figure S3.1:** Elastic vs inelastic processes. We define transmission of the electrons when the scattering angle satisfies the condition $\theta < 90°$ and reflection of the electrons when $\theta > 90°$.

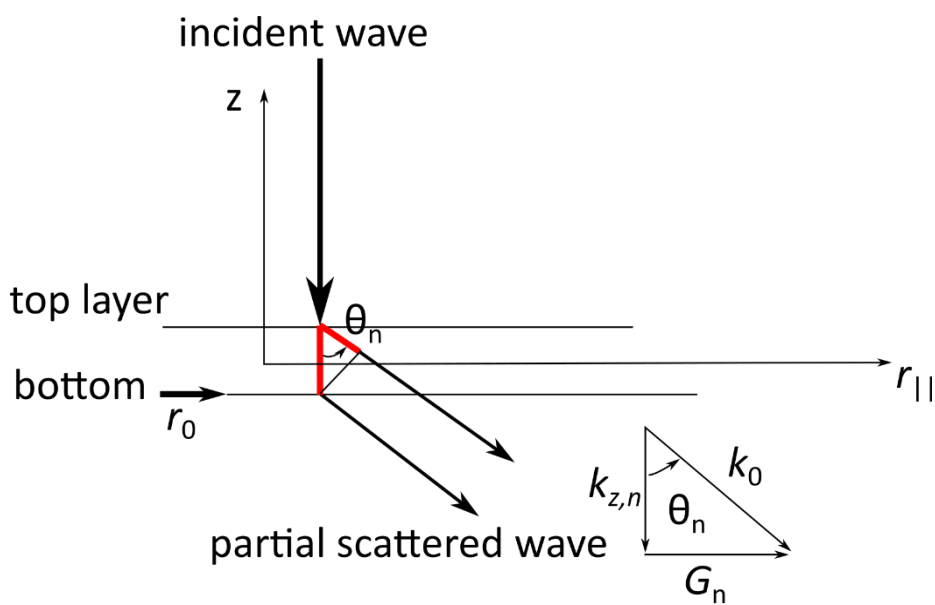

**Figure S3.2:** Geometry of elastic scattering occurring within the two layers of the TMD.

First, we consider modelling of the elastic scattering behaviour, the geometry of which is illustrated in **Figure S3. 2**. Energy conservation for scattering gives:

$$k_0^2 = k_{z,n}^2 + G_n^2 \; ; (1),$$

where $k_0$ is the wave vector of incident electrons, and $G_n$ is a reciprocal vector of the TMD bilayers (not necessarily the basic reciprocal vector). The direction of the partially scattered electron wave ($G_n$) is given by the angle:

$$\theta_n = \arcsin\left(\frac{G_n}{k_0}\right); (2).$$

The partial wave, $\psi_n$, scattered in the angle $\theta_n$, is given by a sum of waves scattered from the top and bottom TMD layers,

$$\psi_n = Ae^{ik_{z,n}z + iG_n r_\parallel} + Ae^{ik_{z,n}z + iG_n(r_\parallel - r_0) + i\Delta\varphi_n}; (3).$$

Here, $\Delta\varphi_n = k_0 d(1 - \cos(\theta_n)) = d(k_0 - k_{z,n}) = d\left(k_0 - \sqrt{k_0^2 - G_n^2}\right)$ is a phase shift coming from the difference in path lengths (red lines in Figure S3.2), and $r_0$ is offset between lattices of the two layers ($r_0 = \pm(0, a/\sqrt{3})$ for MX' (top sign) XM' (bottom sign) stackings and $r_0 = 0$ for XX stacking).

The intensity of the partial wave reaching a detector is,

$$I_n = |\psi_n|^2 = 2A^2(1 + \cos(G_n r_0 - \Delta\varphi_n)); (4).$$

The difference in the intensity for the two domains, $I_{MX} - I_{XM}$ and the sum intensity for the two domains, $I_{MX} + I_{XM}$ are therefore given by:

$$I_{MX} - I_{XM} \propto A(E, \theta_n)\sin(G_n r_0^{MX})\sin(\Delta\varphi_n(E)); (5),$$

$$I_{MX} + I_{XM} \propto A(E, \theta_n)[1 + \cos(G_n r_0^{MX})\cos(\Delta\varphi_n(E))]; (6).$$

The $n$ index subscript is related to the 2D reciprocal vector indexes $\{n_1, n_2\}$ by the following equation (see **Figure S3.3**):

$$G_n = n_1 G_1 + n_2 G_2,$$

$$G_1 = \frac{4\pi}{a\sqrt{3}}\left(-\frac{1}{2}, -\frac{\sqrt{3}}{2}\right), \qquad G_2 = \frac{4\pi}{a\sqrt{3}}(0,1).$$

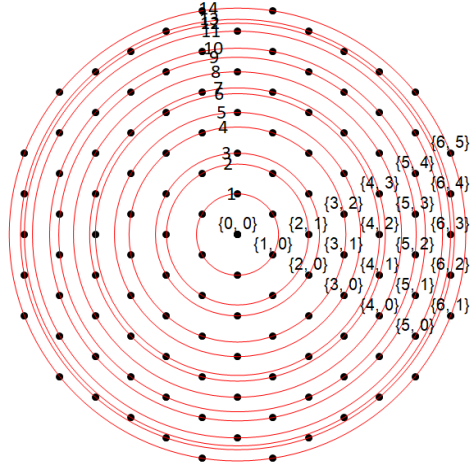

**Figure S3.3:** Relationship between the $n$ index used in the explanation of elastic scattering and the 2D reciprocal vector indexes where $n$ =1 or {1 0}, is equivalent to the Miller Bravais crystallographic notation of {1-100} and where $n$ =2 or {2 1}, is equivalent to the Miller Bravais crystallographic notation of {-21-10} etc.

The scattering angle $\theta_n$ values depend on the accelerating voltage (V) of the primary electron beam as this determines the primary electron energy, $E_n$. The energy dependent factor is given by:

$$A(E, \theta_n) =$$
$$|\langle k_{z,n}, \boldsymbol{k}_{||} = \boldsymbol{G}_n | V_{\text{Mo}}(\boldsymbol{r}_{||}, z) + V_{\text{S}}(\boldsymbol{r}_{||} + \boldsymbol{\tau}, z + d_{XX}/2) + V_{\text{S}}(\boldsymbol{r}_{||} + \boldsymbol{\tau}, z - d_{XX}/2)|k_0, \boldsymbol{k}_{||} = 0\rangle|^2; \quad (7).$$

This takes into account angular dependence of the Bragg scattering probability in the Born approximation with the Thomas-Fermi approximation for potentials ($V_{\text{Mo,S}}$) of molybdenum (Mo) and sulphur (S) atoms; $\boldsymbol{\tau} = (0, a/\sqrt{3})$ and where $\pm d_{XX}/2$ are in-plane and out-of-plane distances between Mo and S atoms in the same layer.

The difference in the intensity for the two domains, $I_{MX} - I_{XM}$ and the sum intensity for the two domains, $I_{MX} + I_{XM}$ as a function of $E_n$ are presented in **Figure S3.4a** and **b**, respectively. Here each line, $\theta_n(E) = \arcsin\left(\frac{G_n}{k_0(E)}\right)$, corresponds to a particular value of $n$ characterised by a particular reciprocal vector $G_n$. Colour maps superimposed on the lines show the magnitude of the intensity contrast between domains and total scattering intensity for **Figure S3.4a** and **b**, respectively.

In **Figure S3.4a**, many scattering angles are seen to give very small or zero intensity for a range of $E_n$. This happens for two reasons: either (i) $I_{MX} - I_{XM} \propto \sin(\boldsymbol{G}_n \boldsymbol{r}_0^{MX}) = 0$ for given set of 6 reciprocal scattering vectors or (ii) there are angles which contain 12 reciprocal scattering vectors, for which the two symmetrically equivalent sets of 6 reciprocal vectors give opposite values for $\sin(\boldsymbol{G}_n \boldsymbol{r}_0^{MX})$ leading to zero total contribution.

**Figure S3.4c** and **Figure S3.4e** show the values of the scattering intensities for a primary electron energy of 1500 eV, providing plots that are similar to those obtained from powder diffraction measurements. **Figure S3.4d** illustrates that at 1500 eV incident beam energy, the significant difference in the intensity of the two domains is for the n=1 ({1-100} Miller Bravais) type reflections.

A closer inspection of the energy dependence of $G_n$ reflections in **Figure S3.4a**, reveals that it is possible to selectively choose the collection angle $\beta$ in order to achieve a reversal of domain contrast, $I_{XM} - I_{MX}$ for primary electron energies of >2000eV and at ~1000eV because for these energies, values of $I_{XM} - I_{MX}$ are positive for at least one $G_n$ and negative for other $G_n$. Contrast reversal is achieved by selecting the appropriate $G_n$ as exemplified in **Figure S3.5**. However, for the energy interval of 900 – 1500 eV, $I_{XM} - I_{MX}$ is negative for all $G_n$ reflections, so it is impossible to invert the contrast between domains by changing collection angle, when considering just elastic scattering. **Figure S3. 6** shows the elastic contribution to the Michelson contrast ($I_{XM} - I_{MX} / I_{MX} + I_{XM}$) of the domains for different ranges of collection angle ($\beta$) at 1500 eV. This simply shows that to achieve high contrast, it is necessary to choose the angular range to include the scattering from the $n$=1 {1-100} reflections.

A similar consideration of the behaviour of $I_{MX} - I_{XM}$ as a function of $E_n$ shows that when summing elastic scattering contributions over a large angular range the domain contrast will be highest in the interval between 1200 – 1700 eV, since here all c⁴ontributions to $I_{MX} - I_{XM}$ have the same sign.

Elastic scattering also occurs at large scattering angles with angles 90°>θ>180° corresponding to reflected electrons. As illustrated in **Figures S3.7** and **S3.8**, for these large scattering angles there are many more $G_n$ reflections which contribute to the contrast and they are more sensitive to the precise electron beam energy. However, the intensity of these high angle elastically scattered electron is four orders of magnitude smaller than the transmitted electrons and therefore the elastic signal reflected in the bilayer is considered negligible in this work.

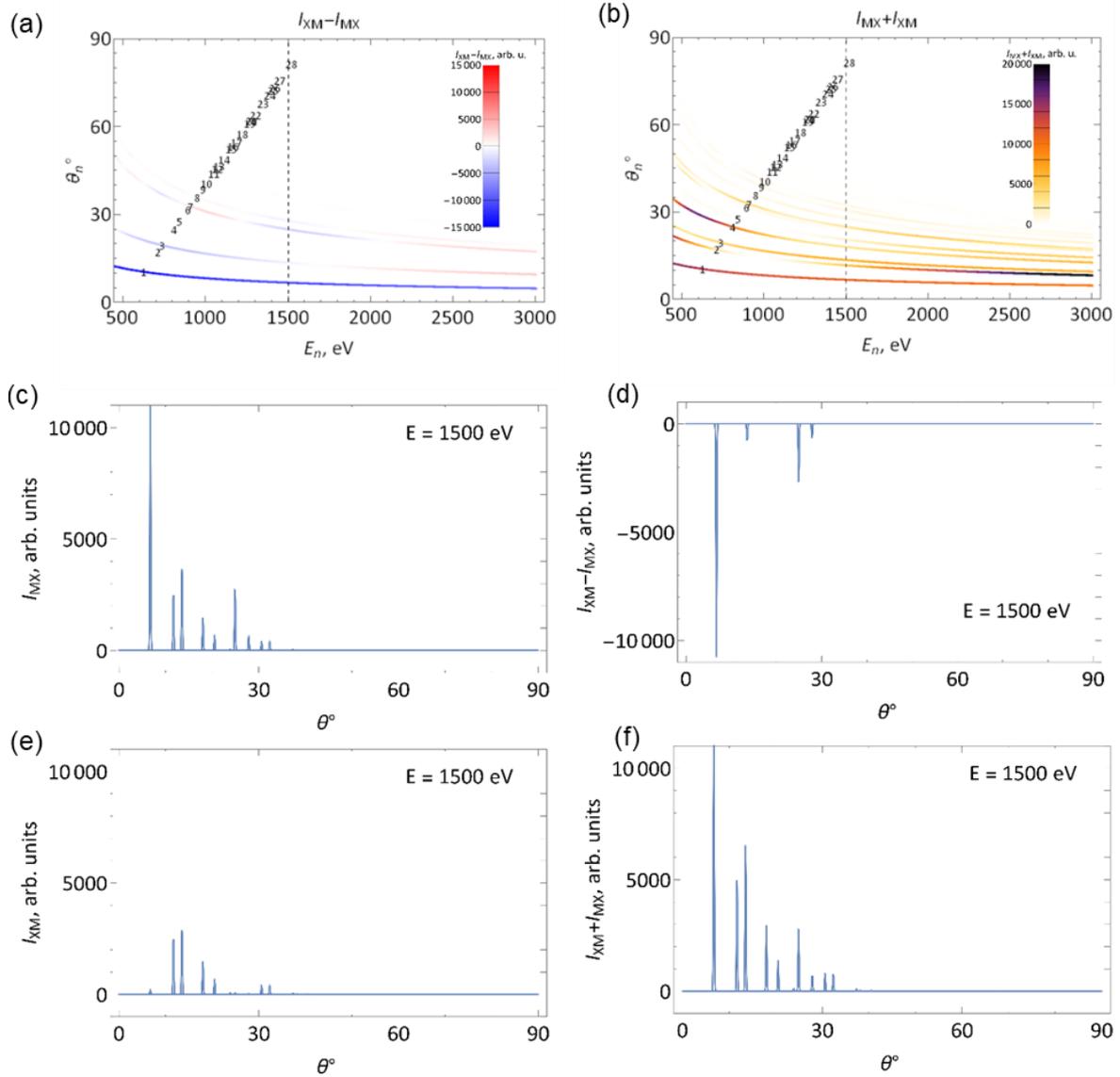

**Figure S3.4: a) & b)**, The difference and sum of the transmitted scattering intensity for the two domains, $I_{XM}$ & $I_{MX}$, both as a function of primary electron beam energy $E_n$ (eV), respectively. This data was used to calculate the theoretical Michaelson contrast for scattering in the range of 0°< θ < 90° (transmission). A colour scale is used to show the scattering intensity where white equates to no intensity. Dashed vertical line highlights the incident electron beam energy of 1500 eV relevant to **c)-f)**. **c) & e)** Intensity of the electron scattering as a function of scattering angle at a primary electron energy of 1500 eV for the MX' and XM' domains, correspondingly. **d) & f)** The difference and the sum of the scattering intensity for the two domains, $I_{XM}$ & $I_{MX}$, both at 1500 eV, respectively.

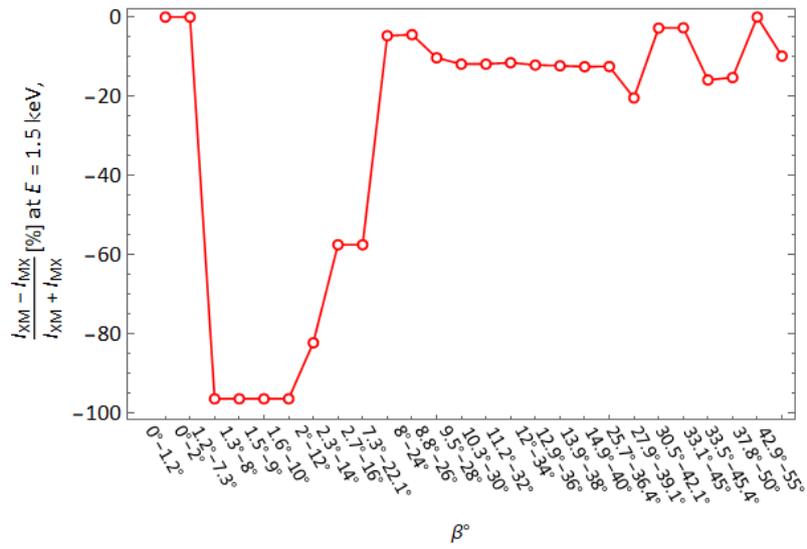

**Figure S3.5:** Summation of the Michaelson contrast for transmitted electrons due to elastic scattering at an incident beam energy of E=1500 eV for different annular intervals of the scattered collection angle.

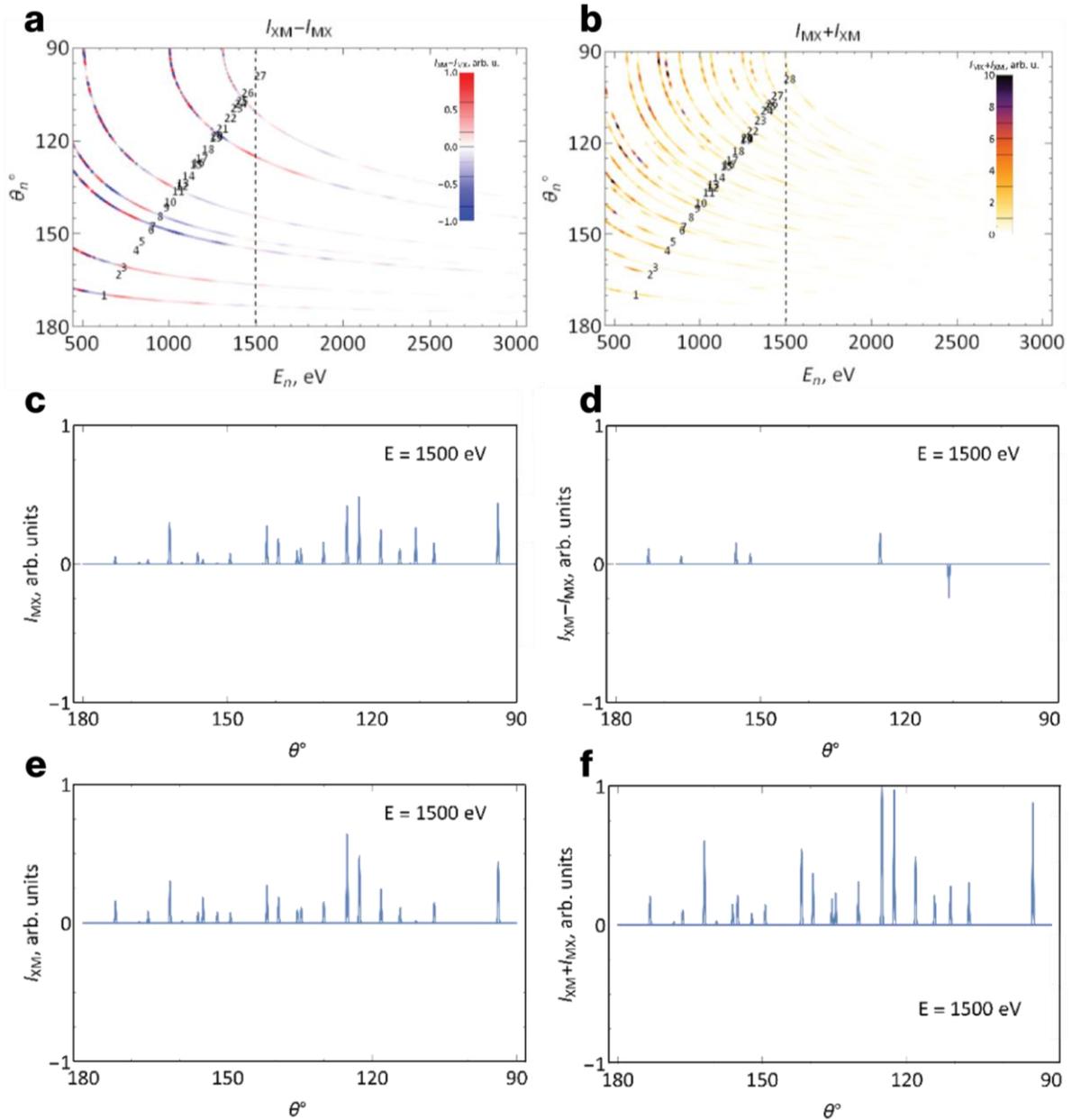

**Figure S3.6: a**) The difference, and **b**) the sum of the reflected scattering intensity for the two domains, $I_{XM}$ & $I_{MX}$, both as a function of primary electron beam energy $E_n$ (eV) used to calculate the theoretical Michaelson contrast for scattering in the range of 90°< θ < 180° (reflection). A colour scale is used to show the scattering intensity where white equates to no intensity. Dashed vertical line highlights the experimental incident electron beam energy of 1500 eV used in this work, at which the intensity plots in **c**)-**f**) were calculated. **c**) & **e**) Intensity of the electron scattering as a function of scattering angle for the MX' and XM' domains, respectively. **d**) & **f**) The difference and the sum of the scattering intensity for the two domains, $I_{XM}$ & $I_{MX}$, respectively. The data is similar to **Figure S3.4** but for reflection elastic processes, the arbitrary units (a.u.) are the same as in **Figure S3.4** demonstrating the magnitude of scattering is ~$10^4$ times lower.

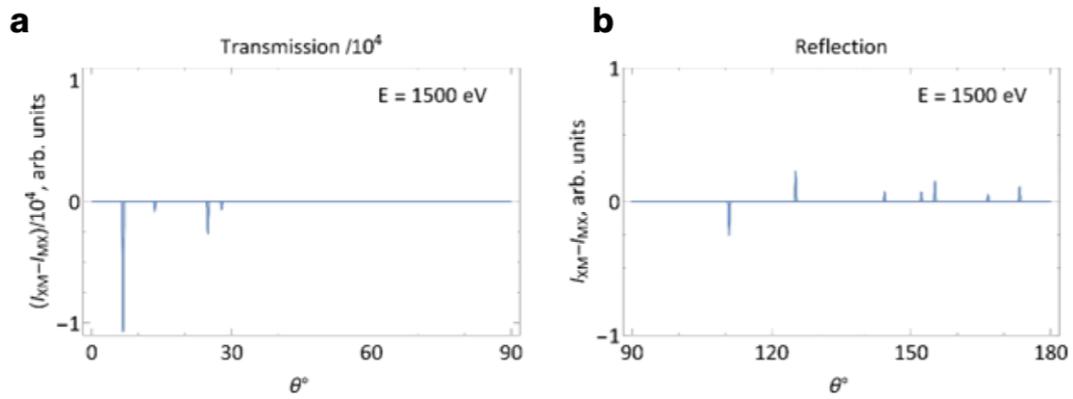

**Figure S3.7:** Comparison of **a**) transmission, and **b**) reflection domain contrast intensities for Bragg scattering at an incident electron beam energy of E=1500 eV. Note that the x-axis is flipped from **Figure S3.6** so theta increases from left to right. Also worth noting is the factor of $10^4$ difference between the transmitted and reflected intensity scales.

## 3.2. Inelastic Scattering

To understand inelastic scattering of high energy incident electrons, we have constructed a simple Monte Carlo model of secondary electron generation and attenuation. This requires understanding of the underlying processes through which primary electrons transfer energy and momentum to bound TMD valence and core electrons, and the associated cross sections and inelastic mean-free paths (IMFPs). These processes fall into two categories, involving inelastic scattering of the primary beam with *valence* and *core* electrons.

### 3.2.1. Inelastic loss processes

The valence electron differential IMFP, $\bar{\mu}(E_0, \omega)$, for a primary beam energy, $E_0$ and energy loss $\omega$, can be calculated from the electron energy loss function, defined as the imaginary part of the inverse dielectric function, $Im\left[-\frac{1}{\epsilon(q,\omega)}\right]$. This quantifies the probability that a primary electron will exchange a given energy and momentum with a materials' valence band, mainly through direct interband transitions and plasmon generation. [3,4] The IMFP is calculated from the energy loss function by integration over all allowed energy and momentum exchanges, $\mu(E_0) = \int_0^E \bar{\mu}(E_0, \omega) \propto \int_{-q}^{q} Im\left[-\frac{1}{\epsilon(q,\omega)}\right]$.

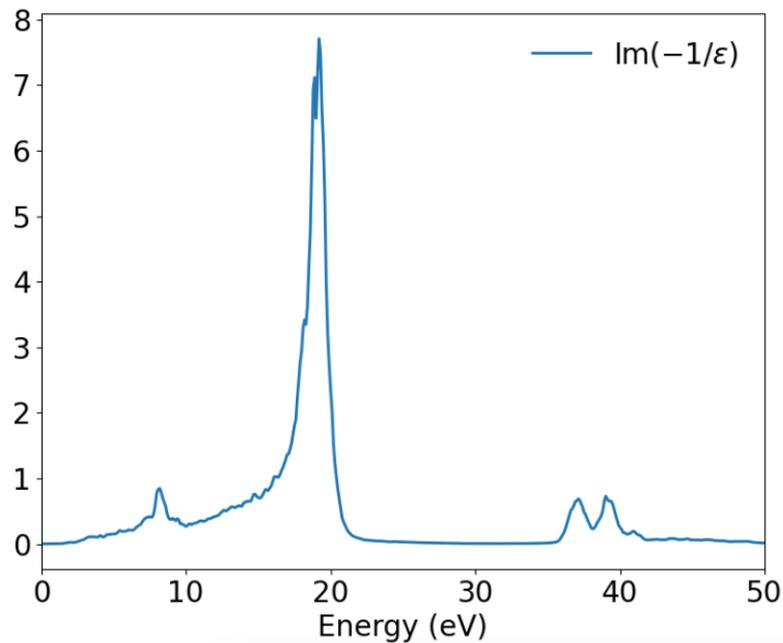

**Figure S3.8**: DFT-calculated energy loss function $Im\left[-\frac{1}{\epsilon(q=0,\omega)}\right]$ for an MoS$_2$ P-bilayer.

To quantify the energy loss spectrum, $Im\left[-\frac{1}{\epsilon(q,\omega)}\right] = \frac{\epsilon_2}{\epsilon_1^2 + \epsilon_2^2}$ was evaluated using density functional theory (DFT) calculations of the real ($\epsilon_1$) and imaginary ($\epsilon_2$) parts of the dielectric function, and is shown in **Figure S3.8**. This shows that valence band inelastic scattering is dominated by small energy losses of the primary beam, with maximum energy loss of around

40 eV from a single collision, and peaks at ~8 eV and ~20 eV which correspond to interband transitions and plasmon loss.[3,4]

For semiconducting materials, such as those considered in this work, the core electron contribution to the IMFP is generally less significant than that from valence electrons, contributing to approximately 10 - 20% of the total IMFP for a primary energy ≤ 1500 eV. The differential IMFP for ionisation by an inner shell electron has a sharp edge at the corresponding binding energy, which for the elements considered here fall in the range of 200-500 eV. Core electron ionisation allows for less frequent, larger energy transfer from the primary beam to the bound electrons.[3,4]

### 3.2.2. Inelastic scattering mean-free paths

Based on these considerations, most electrons generated by scattering by the incident beam will have low energy (< 40 eV), with maximum possible energy transfer of around 500 eV for a single collision (the binding energy of the Mo $M_1$3s orbital, which is the largest sub-1500 eV core electron in MoS$_2$). The energy dependence of the electron-electron IMFP ($\lambda_{mfp}$) in the energy range 50-2000 eV can be calculated using the TPP-2M modification of the Bethe equation[5,6] for energy loss of fast charged particles when passing through matter *via* ionisation.[3,4] This takes the form,

$$\lambda_{mfp} = \frac{E}{E_p^2[\beta \ln \gamma E - C/E + D/E^2]}; \ (8).$$

Here, $\lambda_{mfp}$ is the IMFP in Å, E is electron energy in eV, $E_p = 28.8(N_v\rho/M)$ is the free-electron plasmon energy, $\rho$ is bulk density in gcm$^{-3}$, N$_v$ is the number of valence electrons. M is molecular weight, $E_g$ is the band gap, and $\beta$ and $\gamma$ are additional empirical, material-dependent parameters. Values of $\lambda_{mfp}$ (in Å) for the electron energy range *E* = 500 to 2000 eV, for MoS$_2$, WS$_2$ & hexagonal boron nitride (hBN), using the parameters shown in **Table 1**, are plotted in **Figure S3.9**. The obtained dependence agrees well with available experimental data for the IMFP in MoS$_2$[7,8] and hBN[9,10] in the 50-2000 eV range.

|  | Density, ρ (g/cm³) | No. of valence | Band gap | Free-electron plasmon |
|--|--|--|--|--|

|  | electrons, $N_v$ | energy, $E_g$ (eV) | energy, $E_p$ (eV) |
|---|---|---|---|
| **MoS₂** | 5.06 | 18 | 1.8 | 21.72 |
| **WS₂** | 7.5 | 18 | 2.1 | 21.25 |
| **HBN** | 2.2 | 8 | 5.955 | 24.26 |

**Table 1:** TPP-2M parameters for MoS₂, WS₂ and HBN.

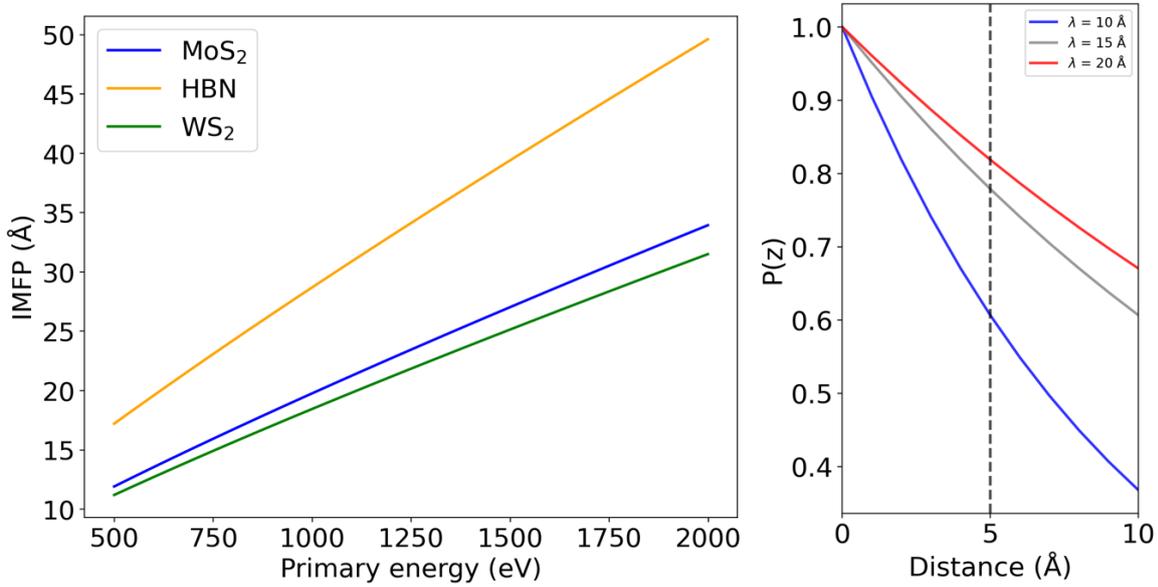

**Figure S3.9:** $\lambda_{mfp}$ (IMFP) vs electron energy for MoS₂, WS₂ & HBN (left). Probability an electron travels a distance $z$, $P(z) = e^{-z/\lambda_{mfp}}$ through an isotropic material with the approximate height of an MoS₂ bilayer for different $\lambda_{mfp}$ in the 500-2000 eV range of MoS₂ (right).

### 3.2.3. Monte Carlo modelling

Based on the above considerations, electrons with energies in the range 25 - 500 eV are most relevant to the measured inelastic signal. Emission of an inelastically scattered electron is considered as a two-step process, where an incident primary electron generates secondary electrons from a point $r'$ within the bilayer with a probability proportional to the local electron density $\propto \rho(r')$ (see **Figure S3.1**), which can then be attenuated by the electron density along the outgoing escape path. The total intensity of secondary electron emission along this path is taken to be a Beer-Lambert style law,

$$I = I_0 \, \sigma_{1.5keV} \int dr' \, \rho(r') \, \exp\left(-\sigma_{low-E} \int ds \, \rho(r' + s[\vec{x} \sin\theta\cos\phi + \vec{y} \sin\theta\sin\phi + \vec{z} \cos\theta])\right); \quad (9).$$

where the first integral in this expression is over the full distribution of three-dimensional electron density (which can be sampled by a Monte Carlo algorithm) and $s$ is the coordinate along the emission path, which is along a direction with a given tilt ($\theta$) and azimuthal ($\phi$) angle.

Low energy attenuation cross sections have been applied to (on average) reproduce the TPP-2M inelastic mean free path for MoS$_2$[11,12] for a given energy (see **Figure S3.10**). Local electron densities are evaluated using the universal Thomas-Fermi equation, which is solved numerically and used to generate the corresponding electron densities for the atomic species Mo and S, $\rho_{Mo}$ and $\rho_S$,[13] which are to create a three-dimensional grid of electron density, $\rho(x, y, z)$, at DFT-relaxed lattice positions of the two-dimensional MoS$_2$ lattice.

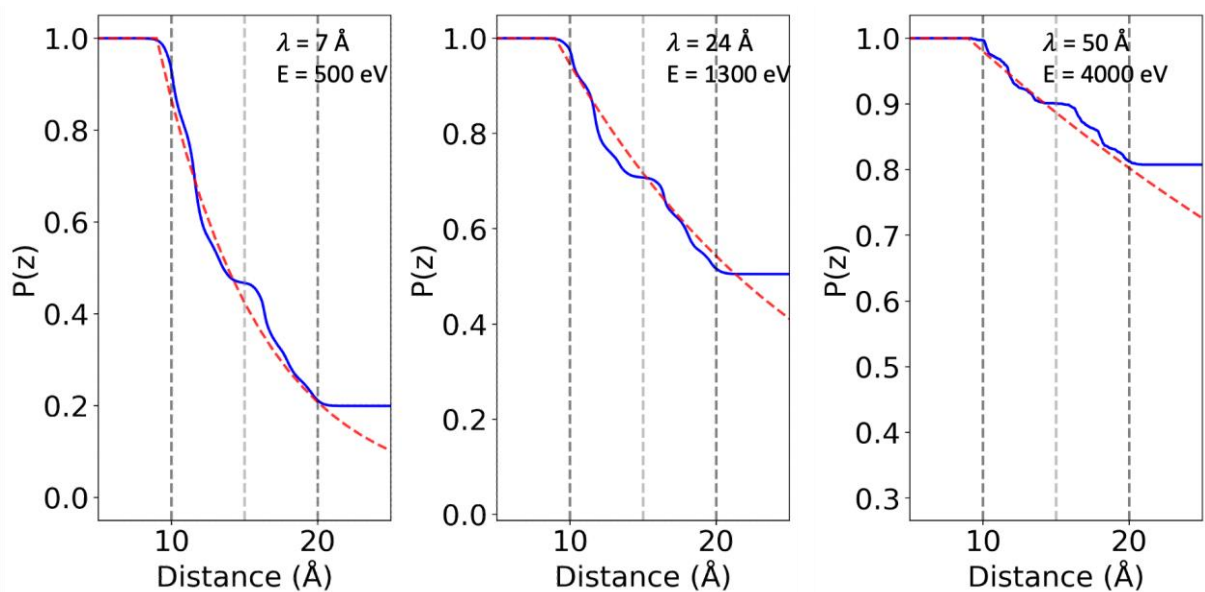

**Figure S3.10**: Probability a primary electron travels a distance $z$, through an MoS$_2$ bilayer, from Monte Carlo simulation of Eq. 9, averaging over 10,000 realisations (blue), and interpolated to the Beer-Lambert exponential dependence, $P(z) = e^{-z/\lambda_{mfp}}$ (red), which enables estimation of $\lambda_{IMFP}$ as a function of electron energy.

Raw emission data ($I_{MX}, I_{XM}$), their difference, $I_{XM} - I_{MX}$ and sum, $I_{XM} + I_{MX}$, as a function of emission scattering angle, are shown in **Figure S3.11**. Emission directions which are along the armchair, $\phi = 30°$, and zigzag $\phi = 60°$ axes (**Figure S3.12**) are referred to as "transmitted" and "reflected" for emission angles at $\theta = 0°$ and 180° to the assumed primary beam direction, respectively. The average contrast, calculated from summing the intensity from both scattering directions is shown in **Figure S3.13**, along with the reflected signal from inelastic electrons generated in the bilayer (a mirror reflection of the forward scattered signal). **Figure S3.14** presents the resulting STEM detector signal (the electron signal from **Figure S3.13** resolved into the relevant experimentally measured annular ranges).

**Figure S3.15** shows the normalised emission intensity as a function of azimuthal angle, $\phi$, from an MX domain for fixed scattering angle $\theta\ from\ 150°\ to\ 160°$, demonstrating a characteristic twist-dependence of emitted secondary electrons. **Figure S3.16** shows the calculated variation in contrast of the inelastic signal for three different secondary electron energies ($E = 25, 100, 800$ eV) using a fixed primary beam energy of $E_0 = 1500$ eV. We observe significantly greater contrast for low energy secondary electrons.

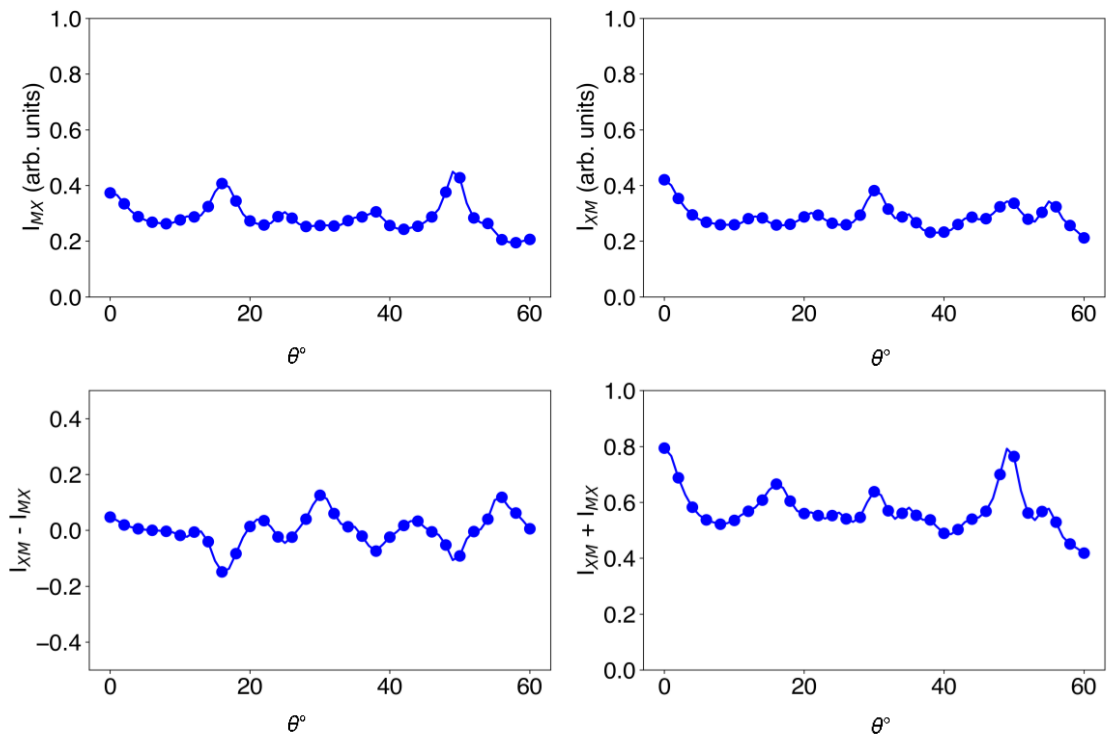

**Figure S3.11:** $I_{XM}$ intensity and $I_{MX}$ intensity, as well as the domain contrast $I_{XM} - I_{MX}$ and sum intensity $I_{XM} + I_{MX}$, for inelastic scattering along the **zigzag** direction ($\phi = 60°$) using an incident electron beam energy of 1500 eV for secondary electrons with E = 100 eV.

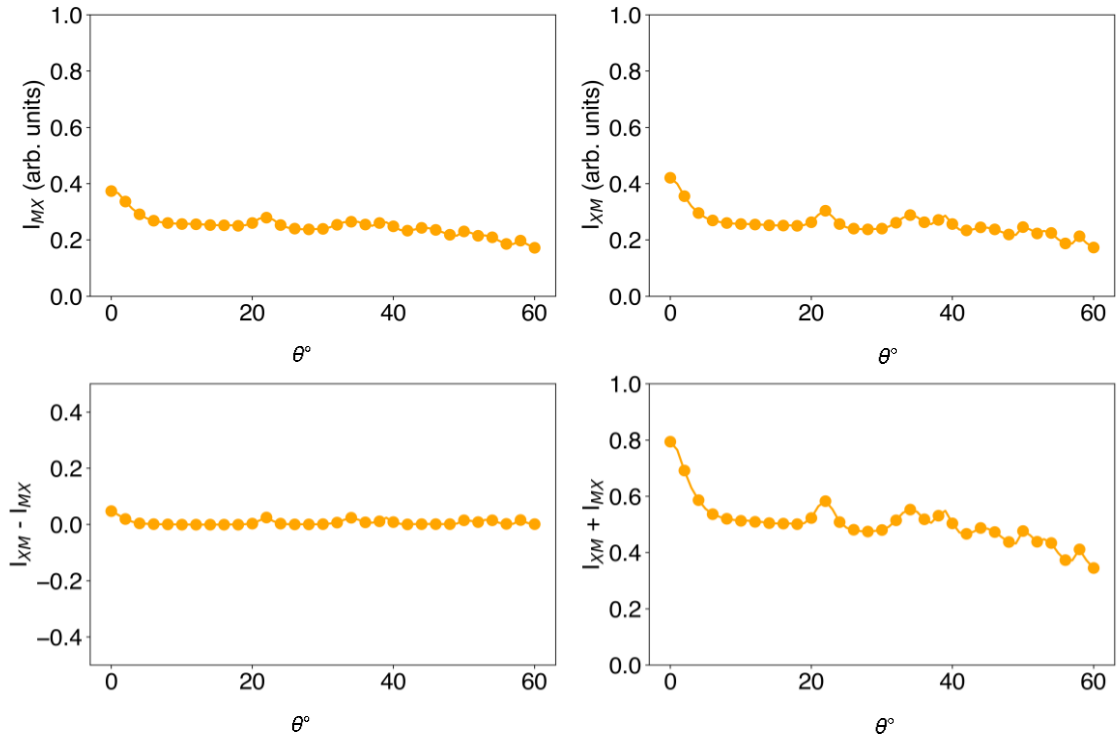

**Figure S3.12**: $I_{XM}$ intensity and $I_{MX}$ intensity, as well as the domain contrast $I_{XM} - I_{MX}$ and sum intensity $I_{XM} + I_{MX}$, for inelastic scattering along the **armchair** direction ($\phi = 30^o$) using an incident electron beam energy of 1500 eV and for secondary electrons with E = 100eV.

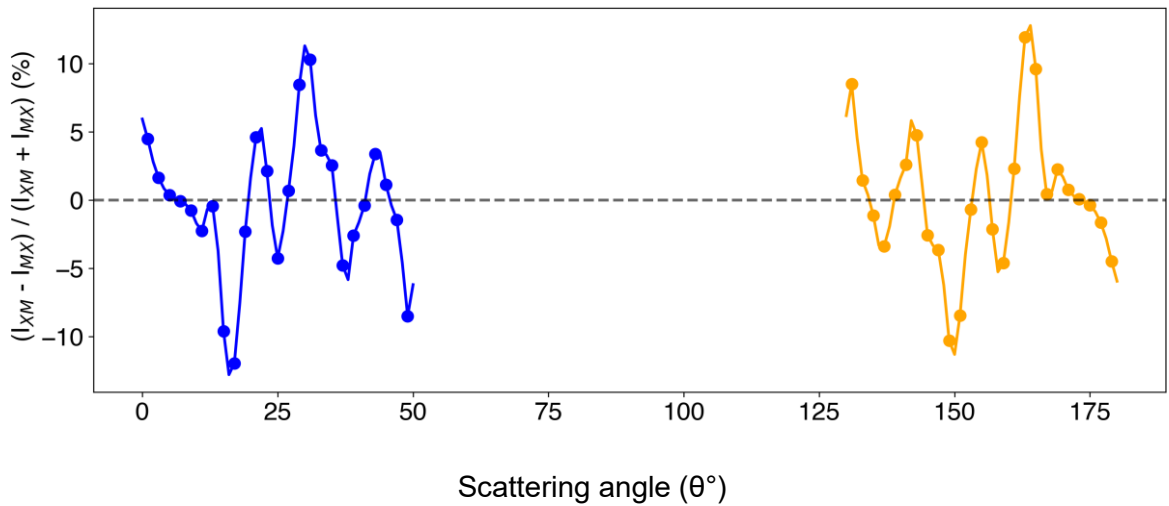

**Figure S3.13**: Theoretical calculations of Michaelson intensity contrast of domains ($I_{XM} - I_{MX} / I_{XM} + I_{MX}$) vs emission angle for transmitted, inelastically scattered electrons (blue), and reflected inelastically scattered electrons (yellow) averaged over both crystallographic directions for generated electrons with E = 100 eV, and a primary beam energy $E_0$ of 1500 eV.

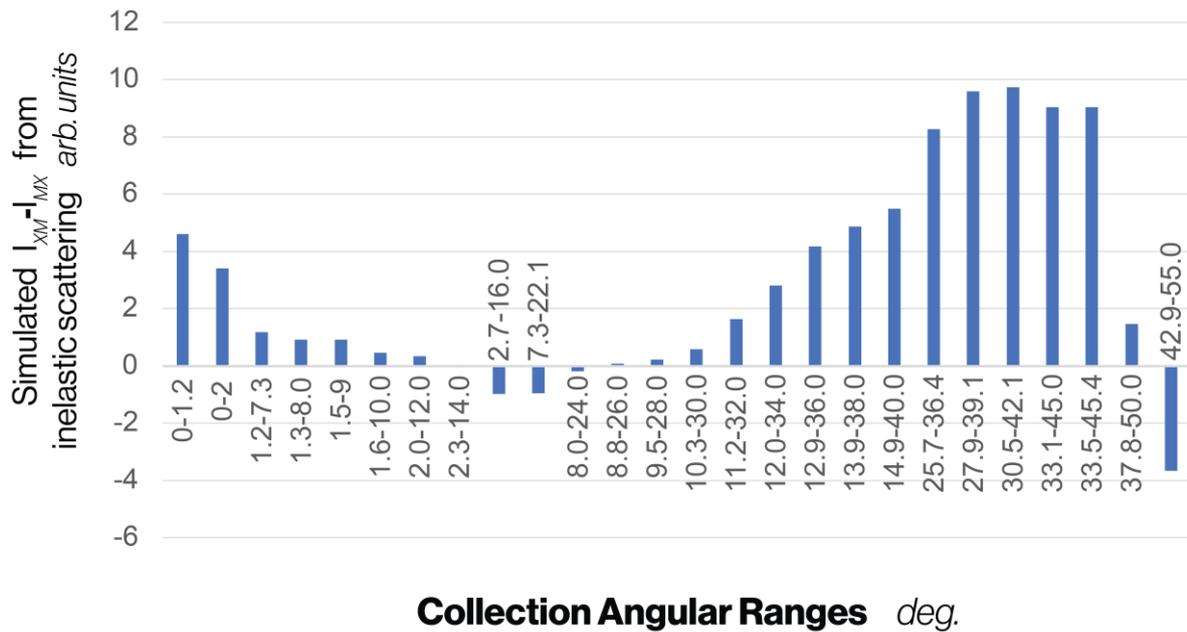

**Figure S3.14:** $I_{XM} - I_{MX}$ (arb units) from the inelastic scattering processes shown as a function of scattering angle for the experimental annular ranges of collection angle (annular range in degrees) for the different STEM detector configurations (similar to Figure S3.5 but for inelastic scattering).

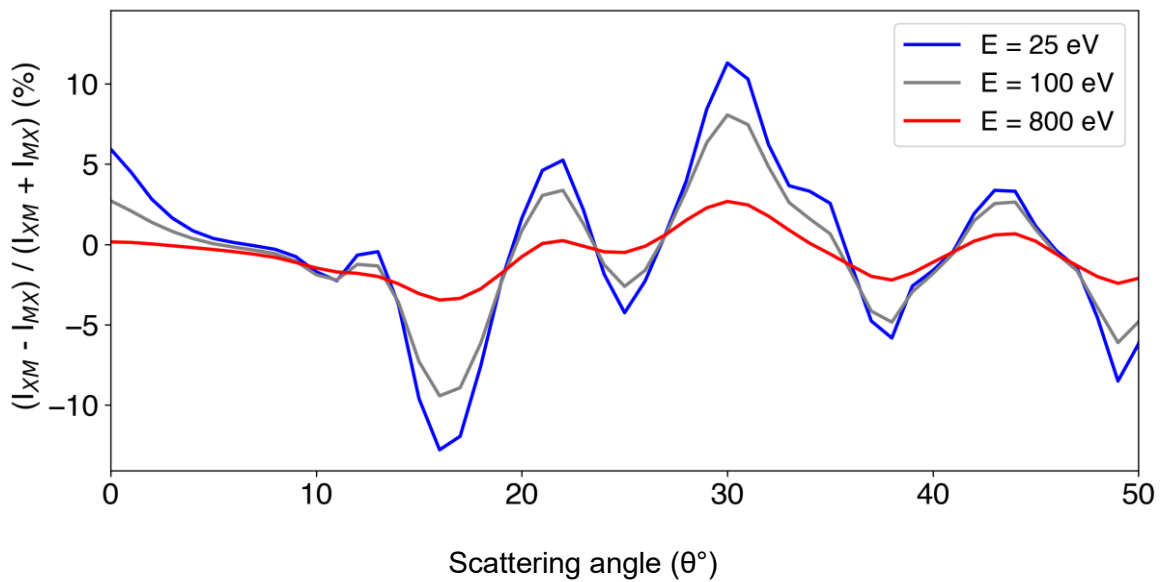

**Figure S3.15**: Azimuthally averaged ($\phi$-averaged) theoretical calculations of Michaelson intensity domain contrast ($I_{XM} - I_{MX} / I_{XM} + I_{MX}$), calculated for a primary beam energy $E_0$ = 1500 eV and secondary electron energies E = 25, 100 and 800 eV. The 100-eV data is the same as that shown in **Figure S3.13**.

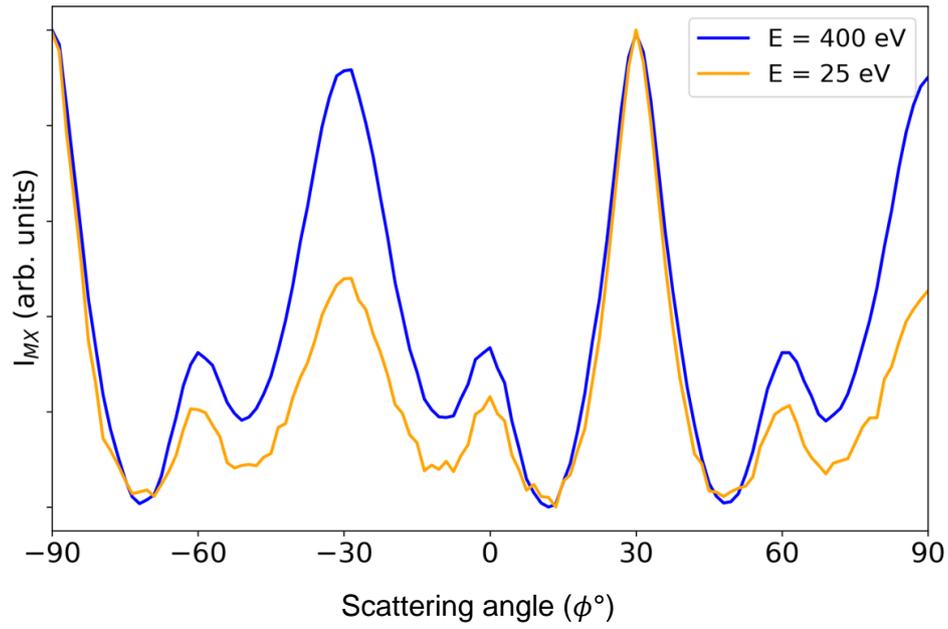

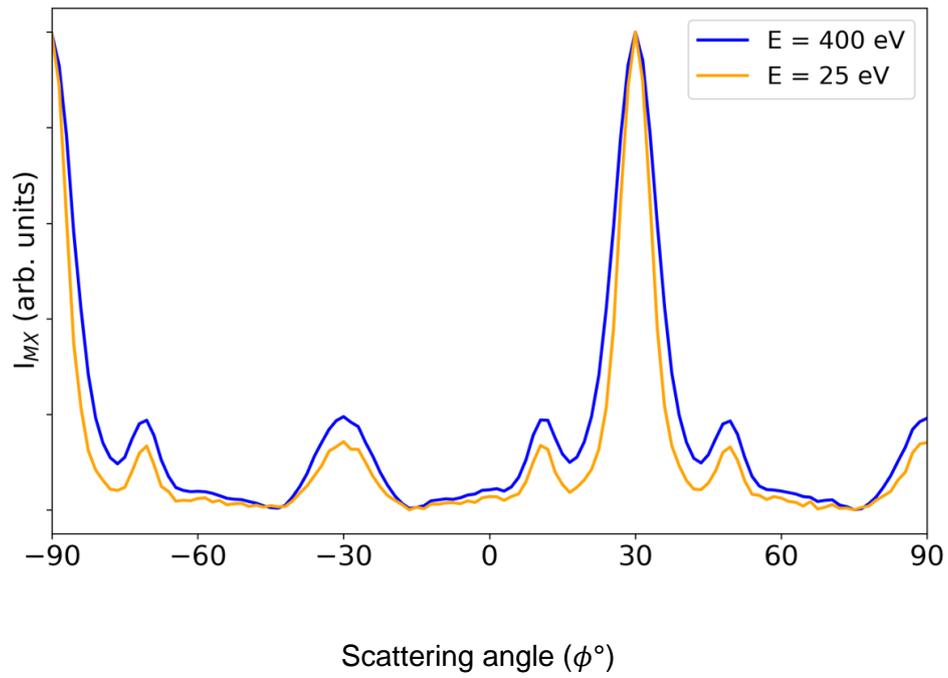

**Figure S3.16**: Emission intensity vs in plane azimuthal angle, $\phi$, for a fixed scattering angle from the transmitted beam of $\theta = 160^o$ (top) and $\theta = 150^o$ (bottom).

## 3.2.4 Angular dependence

Inelastic scattering can also be well-approximated by the double differential cross-section, for an electron of energy $E_0$ to lose energy $\omega$, deflected by an angle θ,

$$\frac{\partial^2 \sigma}{\partial \omega \partial \Omega} \propto \frac{1}{E_0} \frac{1}{\theta^2 + \theta_E^2} \frac{\omega E_P^2 \Delta}{\left[\omega^2 - E_P^2 - 4\gamma E_P E_0 (\theta^2 + \theta_E^2)\right]^2 + \omega^2 \Delta^2}; \textbf{(10)}.$$

where Δ and γ are plasmon damping and dispersion coefficients and the critical angle is related to energy loss as $\theta_E = \omega/E_0$. $\theta_E$ is very small for direct energy loss of the primary beam by valence processes due to the small energy transfers involved.[11,12] **Figure S3.17** shows the angular cross section of a high energy (1500 eV) and low energy (30 eV) electrons scattered by the plasmon peak in MoS$_2$, which shows a very small angular range of generated electrons directly excited by the primary beam.

High-angle scattering will therefore largely arise from second-generation inelastic scattering of secondary electrons, and from high-energy core scattering processes where primary electrons lose a large proportion of their initial kinetic energy, which have a qualitatively similar cross section to the one depicted in **Figure S3.17**. These processes involve multi-generation scattering and core cross-sections which are not considered by the MC model employed here, which we expect becomes less accurate at high energy and large angles.

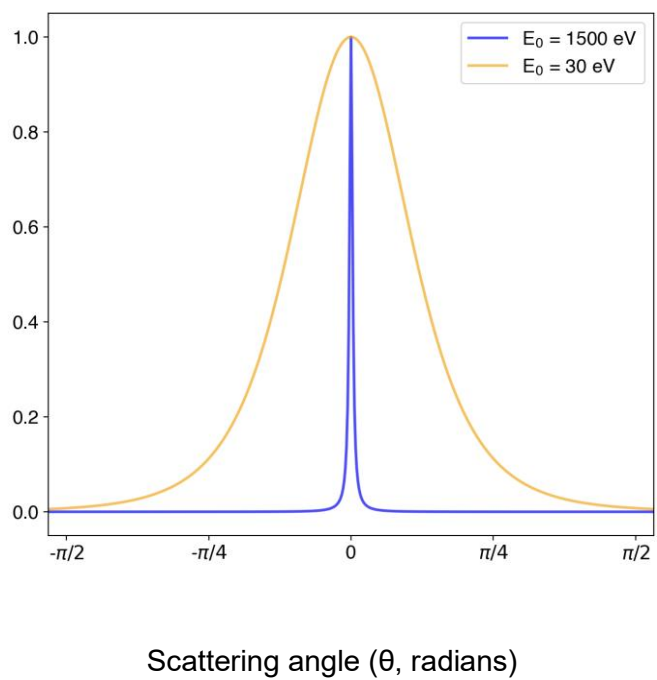

Scattering angle (θ, radians)

**Figure S3.17:** Angular dependence of the double differential cross section, integrated over all allowed energies for plasmon scattering of the primary electron beam.

# 4. Optimising SEM Instrument Parameters

## 4.1 Domain contrast comparisons with varying SEM parameters

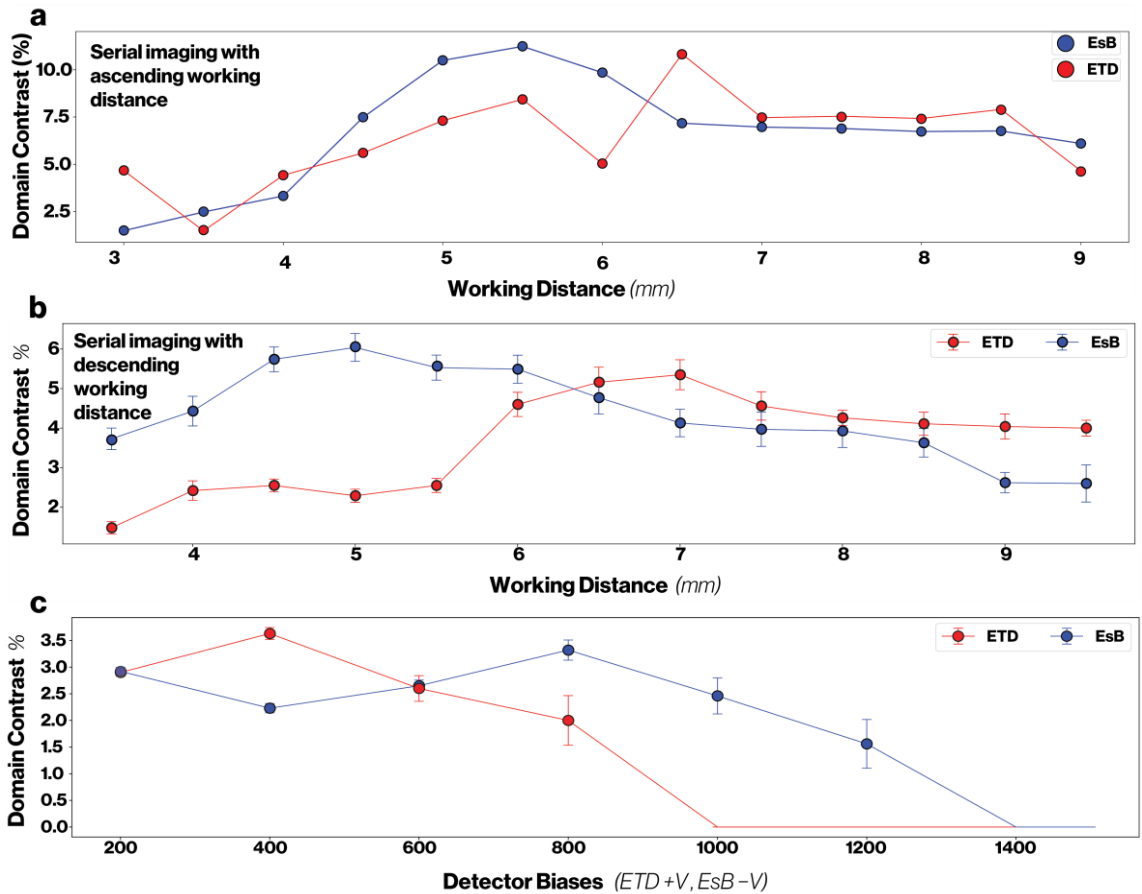

**Figure S4.1:** Effect of **a**) ascending, and **b**) descending working distance on domain contrast using fixed detector biases of -800 V for EsB and +400 V for ETD. **c**) Effect of detector bias on the domain contrast observed from the ETD and EsB signals when imaging a twisted 3R type $MoS_2$ bilayer with a fixed working distance of 4.5 mm. An accelerating voltage of 1.5 kV was used for all data acquisition.

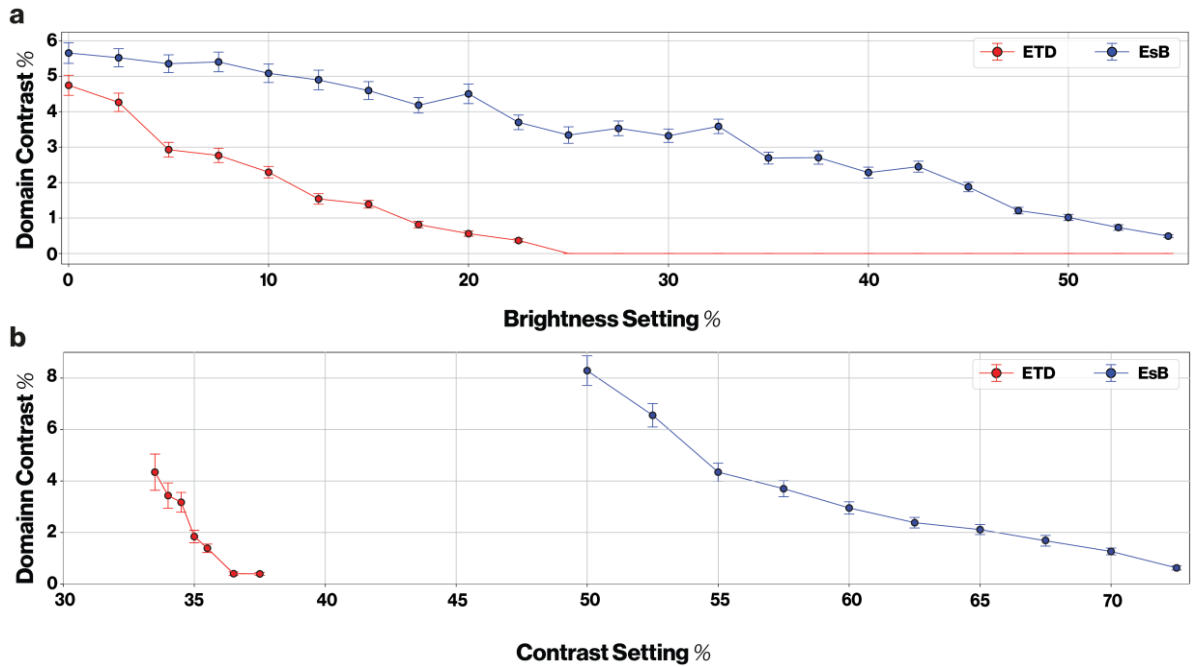

**Figure S4.2: Domain contrast measured with varying brightness and contrast settings of the ETD and EsB detectors**. The EsB domain contrast is found to be more robust to the precise a) brightness, and b) contrast settings, making domains imaged using this detector more readily observed. All Images used for contrast measurement were acquired at a 25 kX magnification, 5 mm working distance, 1.5 kV acceleration voltage, -800 V EsB detector bias and +400V ETD detector bias. The system vacuum was measured to be $1.4 \times 10^{-6}$ mbar.

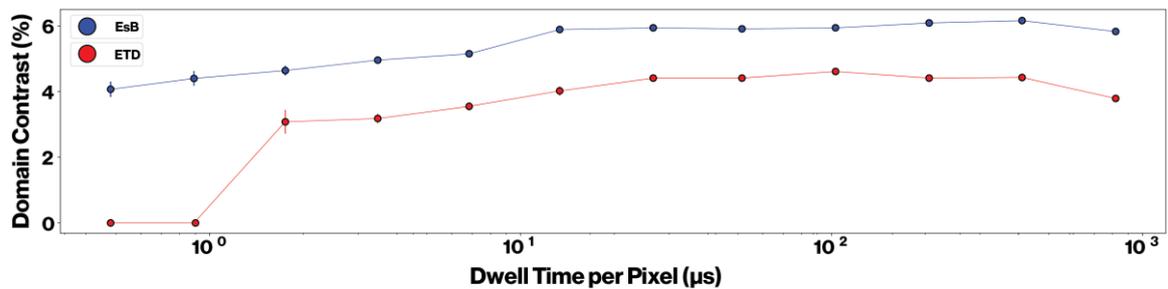

**Figure S4.3: Effect of pixel dwell time on the domain contrast acquired by EsB and ETD detectors.** EsB imaging has a lower minimum dwell time compared to ETD. Images were all acquired at 25 kX magnification, 5.2 mm working distance, 1.5 kV acceleration voltage, -500 V EsB detector bias and +500 V ETD detector bias. Brightness and contrast settings were 0.6% and 31.4% for the ETD, and 18.1% and 54.3% for the EsB, respectively. The system vacuum was measured to be $1.4 \times 10^{-6}$ mbar.

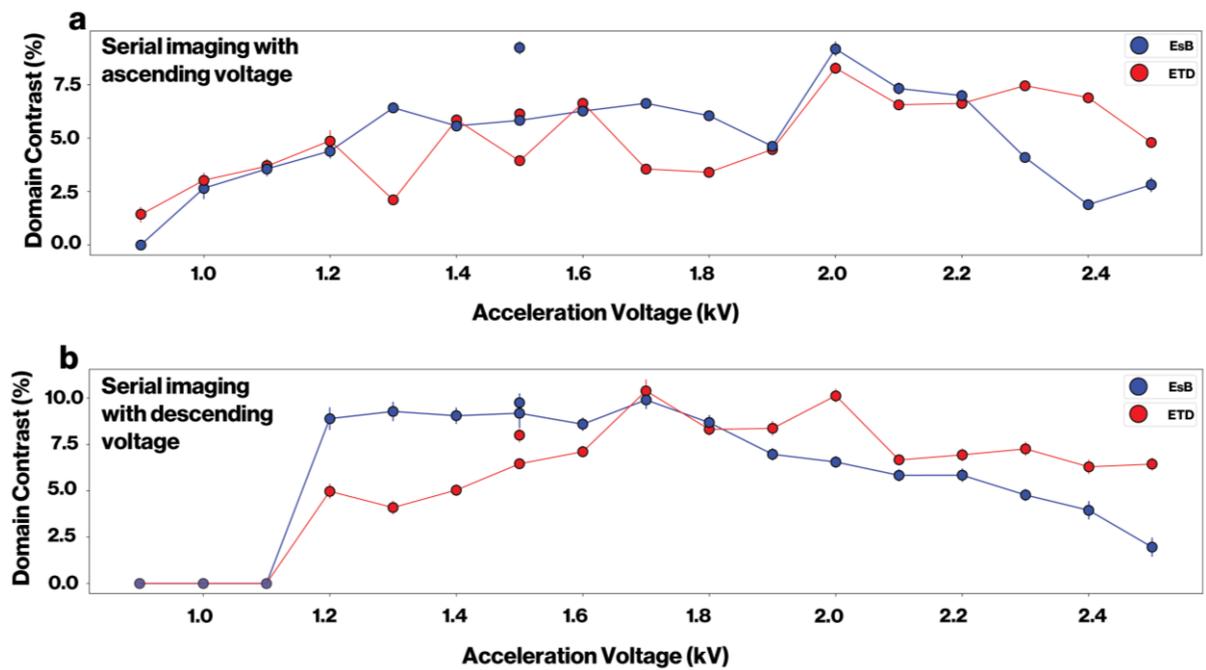

**Figure S4.4: Effect of accelerating voltage on domain contrast acquired by EsB and ETD detectors.** Serial imaging conducted with a) ascending, and b) descending voltages. The separate data points at 1.5 kV were acquired before starting the image series. The lower contrast values measured from the series data at the same accelerating voltage is due to the build-up of surface contamination.

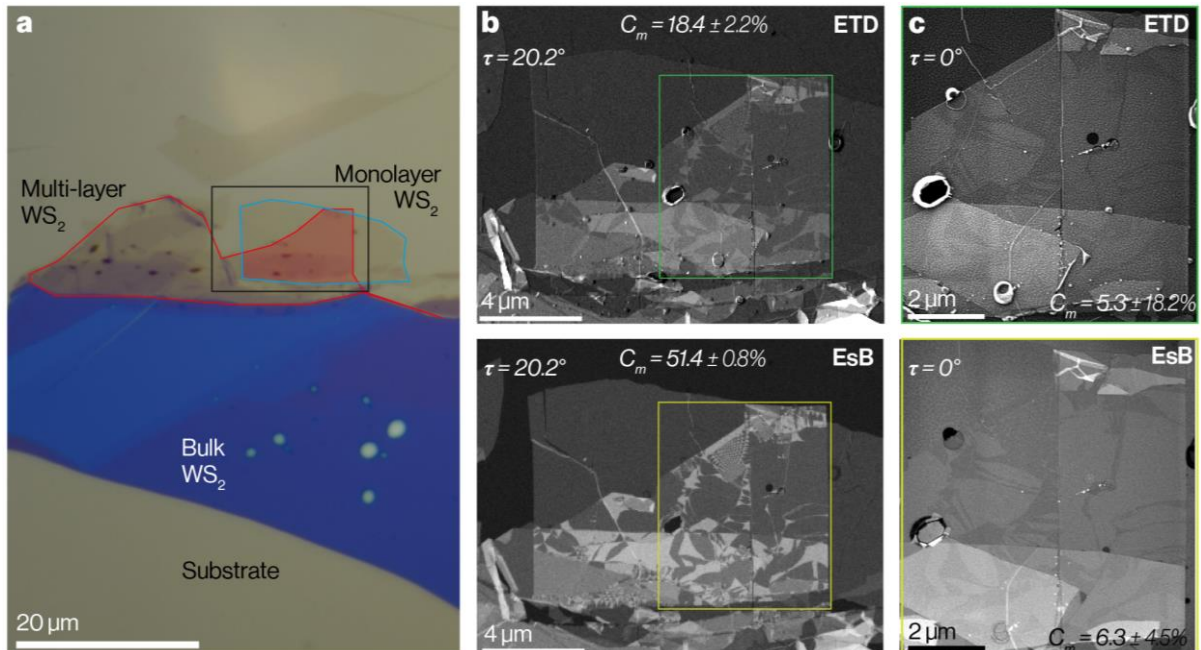

**Figure S4.5: Observed domain contrast change with stage tilt for $WS_2$. a**) Optical micrograph of monolayer $WS_2$ (outlined in blue) placed atop a 1-3 layers thick $WS_2$ flake (outlined in red). **b,c**) SEM images at b) 20.2° and c) 0° stage tilt for ETD (top) and EsB (bottom) signals. The areas shown in c) correspond to the regions highlighted in b). A contrast inversion can be observed due to stage tilt.

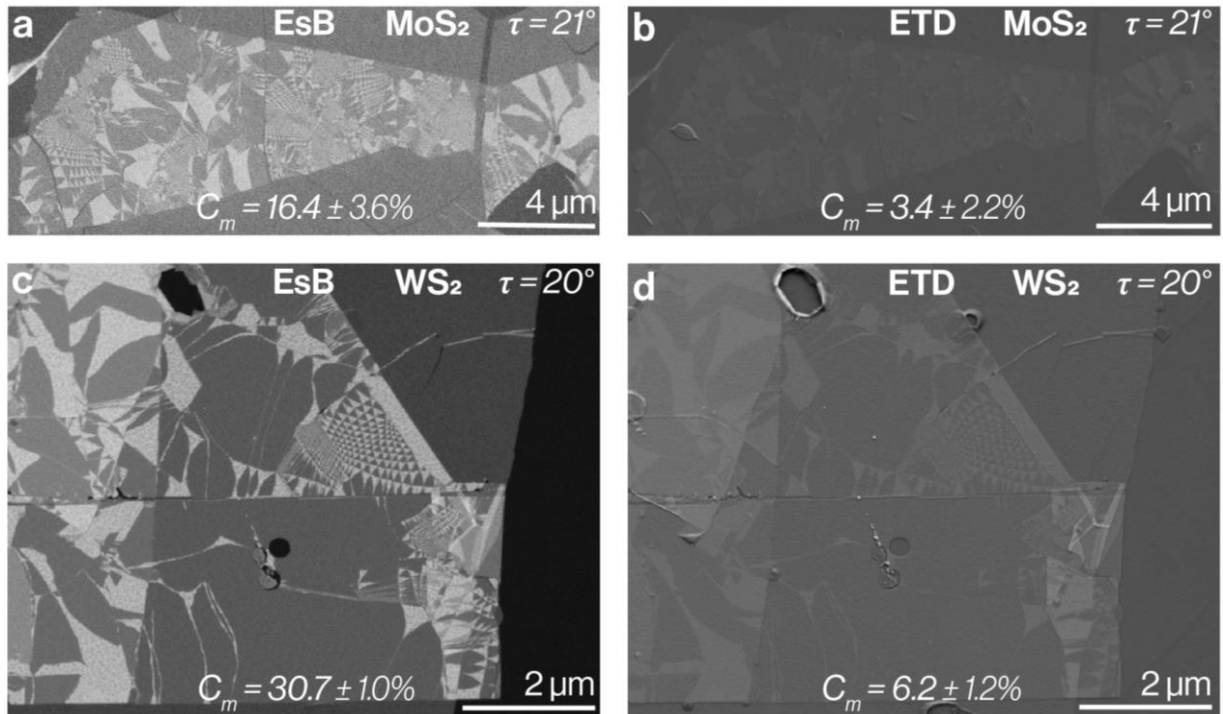

**Figure S4.6: Domain contrast comparison between different TMD bilayers. a,b)** $MoS_2$. imaged with the a) EsB, and b) ETD detectors. **c,d)** $WS_2$. imaged with the c) EsB, and d) ETD detectors. All images were acquired with similar instrumental parameters (WD = 4.2-5.5 mm, acceleration voltage = 1.5 kV, detector biases have been set to -/+ 500 eV for the EsB and ETD electron detectors, respectively).

As the BSE (EsB) detectors are rotationally symmetric, rotation of the sample about the imaging direction, φ, at zero specimen tilt, τ, results in no contrast change (**Figure S4.7**). There is also little change observed from the ETD signal, which is expected as previous work required large specimen tilt to measure azimuthal contrast variations.[14,15]

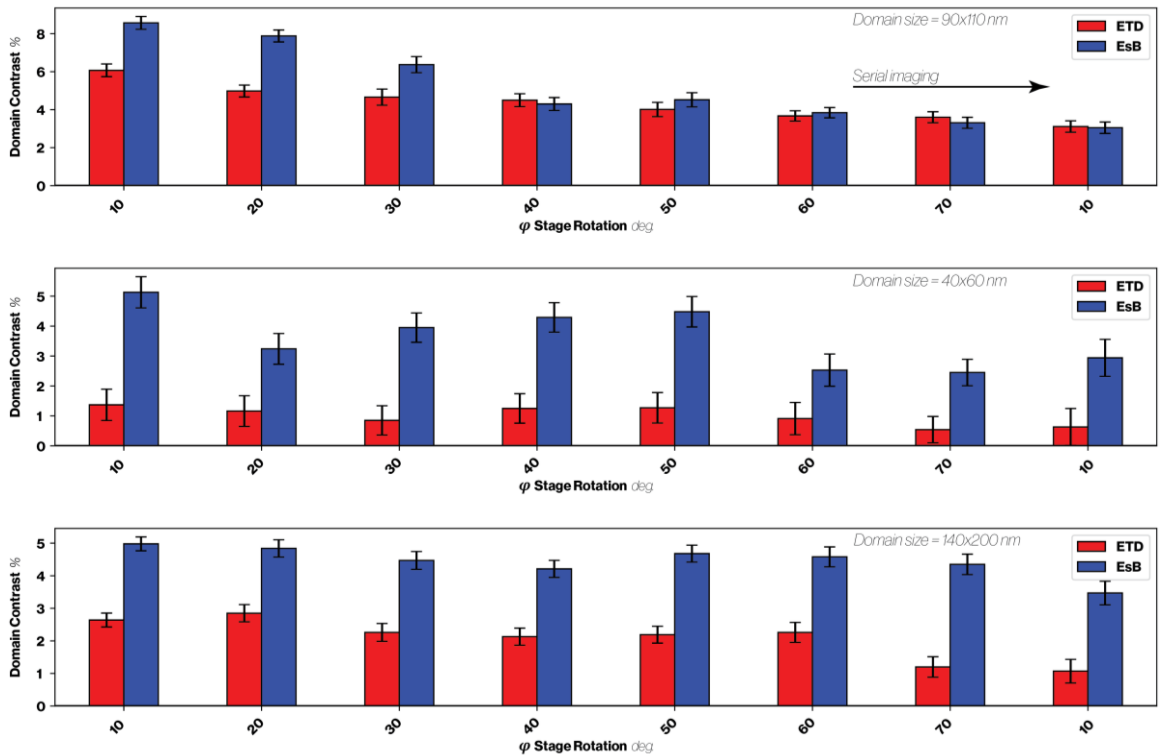

**Figure S4.7: Domain contrast change with stage rotation** in increments of 10° at 0° tilt for three different samples and different imaging areas (given inset). Note that after 7 images at increments of 10° the right-hand side image is of the sample back at 10°. All data sets showed reduced contrast after the image series which is attributed to the build-up of surface contamination, although the extent of contrast reduction during data acquisition depended greatly on the cleanliness of the sample and SEM chamber. Electron images were acquired at 54 kX magnification, 5.3 mm working distance, 1.5 kV acceleration voltage and a beam current of 1 nA.

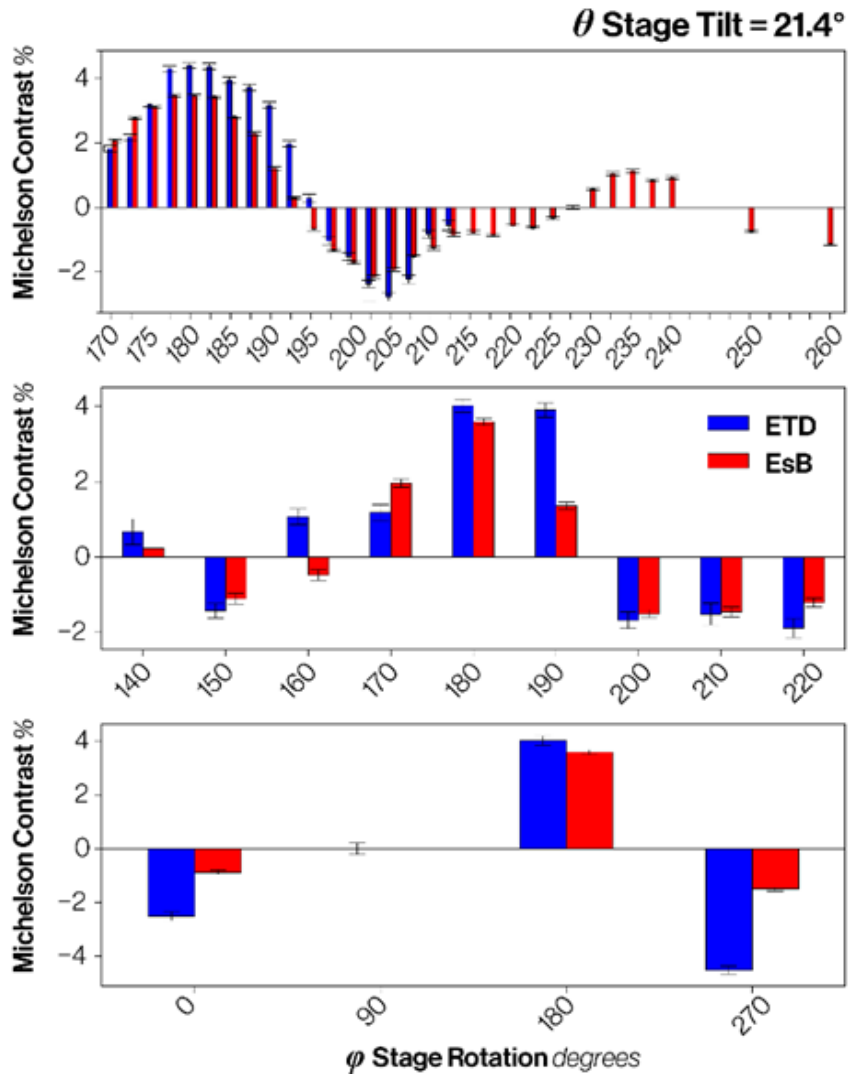

**Figure S4.8**: Domain contrast change with stage rotation 21.4° tilt for three different samples and different imaging areas (given inset). Note that after multiple images all data sets showed reduced contrast which is attributed to the build-up of surface contamination.

## 4.2 Effect of beam induced carbon contamination on domain contrast

If the SEM chamber is not ultra-clean there is a tendency for hydrocarbon contamination to build up on the surface, which acts to degrade image contrast. This hinders sequential measurements of domain contrast as a function of tilt angle (as seen in the top row of **Figure S4.7**, where after 7 images the contrast at 10° is significantly poorer than in the first image). The contrast reduction is similar to that seen for the presence of hBN encapsulation (main text, **Figure 5**) where the EsB and ETD domain contrast values show similar amounts of degradation (proportional to their original intensity). Carbon contamination of the surface post imaging was measured using AFM and found to have a thickness of ~1nm after acquiring the data in **Figure S4.7**.

Interestingly, we find that plasma cleaning the SEM chamber with the sample in-situ provides a means of removing both adventitious surface contamination and surface contamination deposited by previous SEM imaging. Plasma cleaning was done at 20 W with $N_2$ at ~0.3 mbar pressure for 30 minutes using an *in situ* RF Evactron plasma cleaner. We find that this plasma treatment does not remove the contamination directly but facilitates cleaning of the sample with subsequent SEM imaging as shown in the **Supplementary Video S2** and quantified in **Figure S4.9**. SEM imaging after plasma cleaning was observed to restore the original domain contrast in previously imaged areas where contrast has degraded. After 200 images a ~150% improvement in domain contrast was observed for the previously unimaged areas and ~350% improvement was seen in areas that had been contaminated by previous SEM imaging.

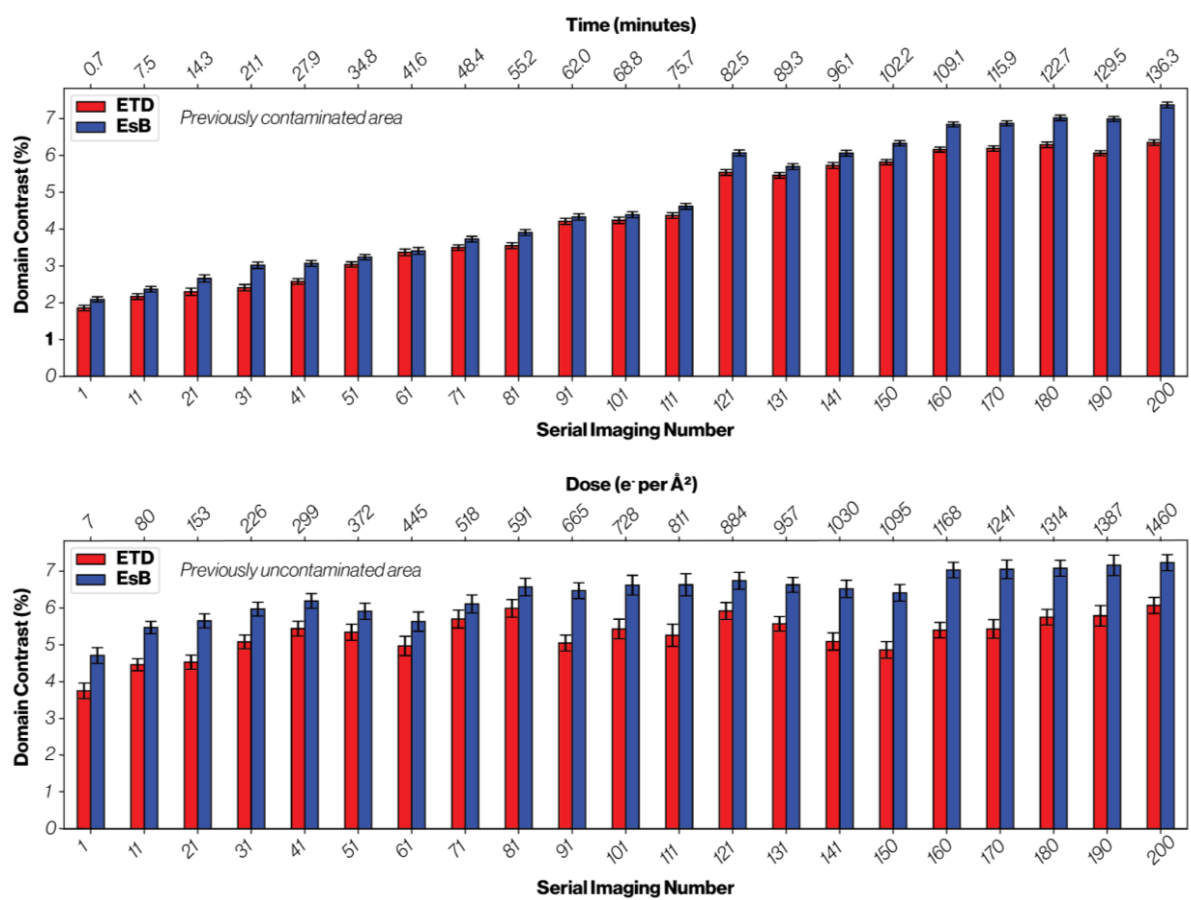

**Figure S4.9: Domain contrast as a function of serial imaging after *in situ* plasma cleaning.** Domain contrast improves with sequential imaging after *in situ* plasma cleaning of the sample as the electron beam removes deposited surface contamination. Video series is shown in Supplementary Movie S3. Top row shows domain contrast for a previously imaged region and bottom row shows domain contrast for a neighbouring region which has not been previously imaged. A dwell time of 52 µs per pixel was used with a probe current of 1 nA and 1.5 kV acceleration voltage.

## 4.3 Imaging MoS₂ domain contrast in different SEMs

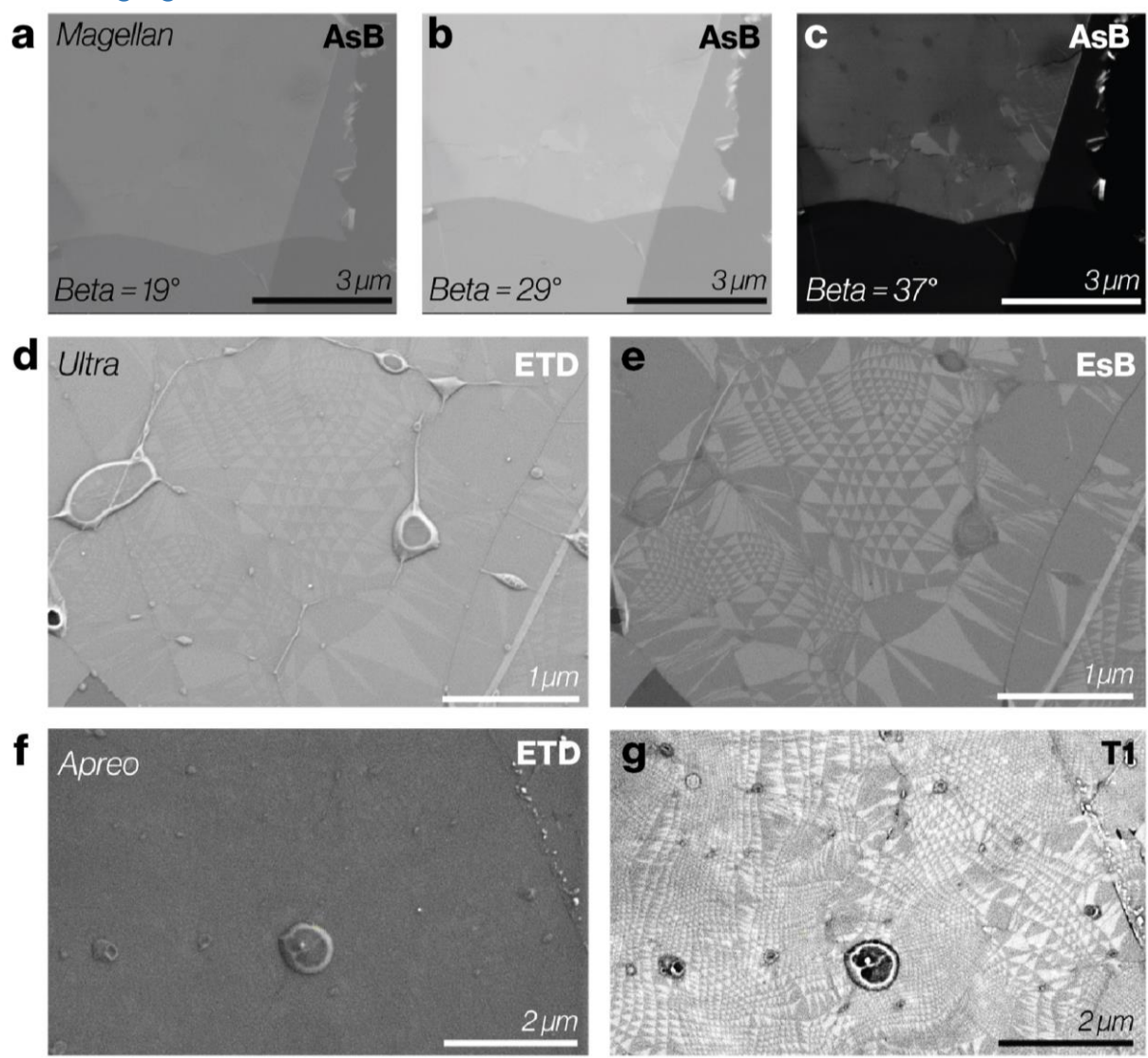

**Figure S4.10: Imaging of twisted domain contrast for MoS$_2$ bilayers using three different SEM columns (all other data presented was obtained on the Zeiss Merlin SEM platform). a-c:** FEI Magellan SEM imaging acquired with the angular selective backscatter (AsB) detector at collection inner angles (beta) of 19°, 29° and 37°, (left to right). **d, e**: Zeiss Ultra SEM imaging comparing ETD and EsB images. **f, g**: FEI Apreo SEM imaging ETD and backscattered electron (T1 detector) images.

## Supplementary Video S1

**Top** ETD and **bottom** EsB electron image tilt series which has been aligned to minimise tilt distortion and drift to highlight changes in domain contrast. Images were acquired at 4.5-8 kX magnification, with a 5.9-6.4 mm working distance and an acceleration voltage of 1.5 kV.

## Supplementary Video S2

**Top** ETD and bottom EsB electron image time series showing contrast improvement during SEM imaging. A poorer contrast region on the left-hand side of the image has resulted from SEM imaging in the presence of carbon contamination. The cleaning (rather than deposition) during imaging results from use of *in situ* plasma cleaning of the sample so that the electron beam sputters away contamination rather than fixing them to the surface.